\documentclass[preprint,aps,superscriptaddress,nofootinbib,showkeys,11pt]{revtex4}
\pdfoutput=1
\usepackage{graphicx}
\usepackage{amsmath}
\usepackage{amsfonts}
\usepackage{amssymb}
\usepackage{xcolor, soul}
\usepackage{epstopdf}
\usepackage{float}
\usepackage{subfigure}
\usepackage{hyperref}
\hypersetup{
     colorlinks   = true,
     citecolor    = red,
     linkcolor    = blue,
     urlcolor     = blue,
}
\def\beq{\begin{equation}}
\def\eeq{\end{equation}}
\def\bea{\begin{eqnarray}}
\def\eea{\end{eqnarray}}
\def\be{\begin{equation}}
\def\ee{\end{equation}}

\def\bse{\begin{subequations}}
\def\ese{\end{subequations}}

\newcommand{\specialcell}[2][c]{\begin{tabular}[#1]{@{}c@{}}#2\end{tabular}}

\graphicspath{{./figs/}}

\begin{document}

\title{Reheating Constraints on Inflaton, Dark Matter: Swampland Conjecture}

\author{Md Riajul Haque}%
\email{riaju176121018@iitg.ac.in}
\author{Debaprasad Maity}
\email{debu@iitg.ac.in}
\affiliation{%
	Department of Physics, Indian Institute of Technology Guwahati.\\
	Guwahati, Assam, India 
}%

\date{\today}

\begin{abstract}
In this paper, we studied the phenomenological consequences of recently proposed swampland conjecture on the inflationary models though constraints on reheating. If dark matter is assumed to be produced during reheating, the conjecture will provide further constraints on the dark matter parameter space through its current relic abundance. As has been pointed out already and also analyzed in our present paper any successful inflationary scenario is in clear tension with the aforementioned conjecture in its current form. However considering the swampland parameters to be free and constrained by the inflationary observables, we studied in detail its consequence on the reheating and dark matter phenomenology. We point out the connection between swampland conjecture and the scalar spectral index $n_s$ by PLANCK within current $2 \sigma$ range, and associated constraints imposed on the reheating temperature and the dark matter annihilation cross-section.

\end{abstract}
\maketitle
\section{Introduction}
Effective field theory framework has been the subject of intensive investigation for its universal appeal to diverse problems in physics. From large scale to small scale where ever there exists a hierarchy of scales in the problem, it proves to be a unique and logical tool to understand the low scale properties by integrating out the high scale modes supplemented with a finite number of scale-dependent free parameters.
However, procedure suggests the existence of theory at a high energy scale, which is in general difficult to define. Therefore, the usual approach is to construct the low energy theory order by order in terms of low energy modes based on some underlying symmetry principle which is assumed to be the full theory property. A natural question then one can ask is whether all possible effective field theory so constructed can have its ultraviolet completion. This is a very difficult question to answer.
String theory has been proved to be a fantastic playing field in this regard. This is the only theory, we know, which is at least an ultraviolet complete theory of gravity.  
  
Recently motivated by this question and taking help of various string theory constructions, a number of attempts have been made to put some constraints on the effective theory which will have consistent VU completion. One such proposal is the swampland conjecture \cite{Ooguri:2018wrx}, which has recently gained interest in the literature. The conjecture says that a low energy effective theory of scalar field minimally coupled with gravity must satisfy the following  universal bound on its form of the potential,
  \bea
\frac{|\nabla V|}{V} \geq \frac {c}{M_p} ,
  \eea
  where $c$ is a dimensionless constant with the magnitude of order unity and $M_{p}$ is the Planck mass.
  However, there exists a refined version of the aforementioned swampland conjecture stated in \cite{Ooguri:2018wrx} which is expressed as,
    \bea \label{swamp1}
   \frac{|\nabla V|}{V} \geq \frac {c}{M_p} ~~~\mbox{or}~~~ ~min(\nabla_i \nabla_j V) \leq -\frac{c'}{M_p^2} . 
    \eea
This is a weaker condition on the possible form of the potential. $c'$ is another universal constant of  order unity, and $min(\nabla_i \nabla_j V)$
   is the minimum eigenvalue of the Hessian of $\nabla_i \nabla_j V$ matrix in an 
   orthogonal frame. A large number of studies have been performed over the years to understand more on the theoretical understanding of this conjecture \cite{Andriot:2018wzk}-\cite{Obied:2018sgi}.
   However, it would be important to mention the interesting debates going on in the literature on the existence of di-Sitter vacuum in sting theory \cite{ds-kklt}\cite{renata}. From the phenomenological point of view, the implication of this conjecture has been widely studied in the context of cosmology \cite{Kehagias:2018uem}\cite{Fukuda:2018haz}\cite{Raveri:2018ddi}. But the main problem  to connect this conjecture with the reheating era is the thermalization portion. However in \cite{Kamali:2019hgv}, author introduce  warm inflation to solve this problem. Starting from inflation to dark energy, the scalar field is ubiquitous and therefore, the conjecture can naturally put constraints on the model building. More interestingly, the hope is that the inflationary, dark energy observation may shed light on UV physics through this conjecture. 
   
   In this paper, we will consider inflationary models with a specific interest on the reheating dynamics. We ask the following question: {\em How does the swampland conjecture put constraints  in the reheating dynamics and the dark matter phenomenology?}. In our analysis, we will consider $(c,c')$ as free parameters. Taking constraints on those parameters from the inflationary dynamics, we will further study the reheating phase.   

In the subsequent section we first  briefly review the basic equations describing the constraints on reheating and consequently on the dark matter parameters considering the CMB anisotropy and the current dark matter abundance. We take four different types of inflationary model potentials and describe how the swampland conjecture restricts the reheating and the dark matter parameter space through the inflationary observables.   
   
   \section{Reheating and dark matter: Methodology} \label{method}
 Reheating is the phase which connects the inflation and big-bang through explosive particle production.
 This phase also can play important role in the dark matter phenomenology. Even though inflation is severely constrained by a large number of cosmological observations, the reheating phase is generally unconstrained. This phase is  parametrized by two important parameters called reheating temperature $T_{re}$ and e-folding number $N_{re}$.
 To go beyond we further incorporate a possible dark matter candidate indirectly originating from the decaying inflaton. During reheating inflaton decays to radiation and then it annihilates to dark matter such that the process during reheating gives us correct relic abundance. As emphasized before, given the observational constraints on the inflationary dynamics, our goal of this paper would be to constrain the reheating and dark matter parameter space through the inflationary parameters considering the swampland conjecture.\\
 For simplicity, we first follow the usual  reheating constraint analysis \cite{Dai:2014jja} where the reheating parameters are calculated assuming the instantaneous conversion of inflaton energy into radiation at the instant of reheating. The evolution during reheating is parametrized by an effective constant equation of state $w_{re}$. 
 Following the assumptions and considering a particular inflation model, one can easily compute the reheating temperature to be 
   \bea \label{eqtre}
T_{re}= \left(\frac{43}{11 g_{re}}\right)^{\frac 1 3}\left(\frac{a_0T_0}{k}\right) H_k e^{-N_k} e^{-N_{re}}~~,
\eea
where $g_{re}$ is the effective number of relativistic degrees of freedom at the instant of reheating. $(T_0= 2.725 K,a_0)$ are the CMB temperature, and the cosmological scale factor at the present time. The number of reheating e-folding number during reheating can be expressed as \cite{Cook:2015vqa}
  \bea
 N_{re}= \frac{4}{(1-3\omega_{re})} \left [-\frac{1}{4} ln \left(\frac{45}{\pi^2 g_{re}}\right) - \frac{1}{3} ln \left(\frac{11 g_{re}}{43}\right)-ln \left(\frac{k}{a_0 T_0}\right)-ln \left(\frac{V_{end}^{1/4}}{H_k}\right)-N_k \right] ~.
\eea 
From the above equation, we can clearly see the appearance of inflationary parameters which are constrained by the swampland conjecture. Therefore, indirect constraints can be imposed on the reheating parameter space. We will be considering some simple canonical scalar field models of inflation, for which Hubble constant $H_k$ and the inflationary e-folding number, $N_k$ are defined as
  \begin{equation}
 H_k= \frac{\pi M_p\sqrt{r_k A_s(n_s^k)}}{\sqrt{2}}~~;~~ N_k= \int_{\phi_{k}}^{\phi_{end}} \frac{|d\phi|}{\sqrt{2\epsilon_V} M_p}~.
\end{equation}
Where, one clearly sees the non-trivial dependence of reheating parameters on the scalar spectral index $n_s^k$, and tensor to scalar ratio $r_k$ through power spectrum of inflaton fluctuation $A_s(n_s^k)$. The aforementioned inflationary parameters, 
\begin{equation}
 n_{s}^{k}= 1- 6 \epsilon_V+ 2 \eta_V~~;~~r_k=16\epsilon_V~,
\end{equation}
in turn depend on the slow roll parameters related to the inflaton potential $V(\phi)$, which can now be constrained by the swampland conjecture, 
\begin{equation}
 \epsilon_V= \frac{M_{p}^2}{2}\left(\frac{V'}{V}\right)^2~~;~~|\eta_V|  = M_{p}^2 \frac{|V''|}{V}~.
\end{equation}
Most importantly all the above quantities are defined at a particular cosmological scale $k$. For CMB, we consider the pivot scale of PLANCK ${k}/{a_0}= 0.002 ~\mbox{Mpc}^{-1}$. 
The end of inflation set the initial condition for the reheating dynamics. Therefore, dynamics will be mostly controlled by $V(\phi_{end})$, where, $\phi_{end}$ is the inflaton field value at the end of inflation follows from the equation $\epsilon_V(\phi_{end}) =1$.

In the discussion so far we have not considered explicit decay of inflaton. However to shed light on the dark matter phenomenology we consider perturbative reheating process where inflation decays to radiation and then radiation to dark matter \cite{Maity:2018dgy}. For this we will have three parameters, the inflaton decay constant $\Gamma_{\phi}$, thermal average of dark matter annihilation cross section $\langle \sigma v\rangle$, and the dark matter mass $M_X$. 
 In the perturbative reheating process, the dynamics of the inflaton energy density $(\rho_\phi)$, the radiation energy density $(\rho_R)$ and dark matter
  particle number density $(n_X)$ are modeled by following homogeneous Boltzmann equations \cite{Giudice:2000ex}.
   \bea
   {\Phi'} &=& -c_1 \frac{A^{1/2}\Phi}{\sqrt{\Phi+ R/A+ X \langle E_X\rangle/m_\phi}}~~;\\
     {R'} &=& c_1 \frac{A^{3/2}\Phi}{\sqrt{\Phi
   		+ R/A+ X \langle E_X\rangle/m_\phi}} + c_2 \frac{A^{-3/2}\langle \sigma v\rangle 2\langle E_X\rangle M_{pl} }{\sqrt{\Phi+ R/A+ X \langle E_X\rangle/m_\phi}}\left(X^2-X_{eq}^2\right)~~;\\ 
      {X'}&=& - c_2 \frac{A^{-5/2}\langle \sigma v\rangle M_{pl} m_\phi }{\sqrt{\Phi+ R/A+ X \langle E_X\rangle/m_\phi}}\left(X^2-X_{eq}^2\right)~~,
   \eea
where, for numerical purpose, new dimensionless variables are defined as
  \bea
  \Phi\equiv\frac{\rho_\phi a^3}{m_\phi}~;~R\equiv \rho_R a^4~;~X\equiv n_X a^3~~.
  \eea
We also rescale the cosmological scale factor by $A=\frac{a}{a_{end}}$, where $a_{end}$ is the scale factor at the end of inflation. 
Prime $'$ represents the derivative with respect to $A$. The inflation mass is $m_{\phi}$. We 
 assume that each $X$ has energy $\langle E_X \rangle=\frac{\rho_X}{n_X} \simeq \sqrt{M^2 + 9 T^2}$. The equilibrium number density of dark matter particle of mass $M_X$  can be expressed in terms of 
modified Bessel function of the second kind:
\bea
n_X^{eq}= \frac{g T^3}{2 \pi^2} \left(\frac{M_X}{T}\right)^2 K_2 \left(\frac{M_X}{T}\right)~.
\eea
\\
 The constants $c_1$ and $c_2$ are defined as,
 \begin{equation}
  c_1= \frac{\sqrt{\frac{3}{8\pi}} M_{pl} \Gamma_\phi}{m_{\phi}^2}~;~c_2=\sqrt{\frac{3}{8\pi}}~~.
  \end{equation}
The initial conditions to solve the  Boltzmann equations are set at the end of reheating to be,
 \bea
 \Phi(1)=\frac{3}{2}\frac{V(\phi_{end})}{m_\phi^4}~;~R(1)=X(1)=0~~.
 \eea
 In this process, we will define reheating temperature ($T_{re}$) from the radiation temperature ($T_{rad}$), at the instant of maximum energy transfer from inflaton to radiation for  $H(t)=\Gamma_\phi$. 
 \bea \label{reheating2}
T_{re}= T_{rad}^{end}= \left(\frac{30}{\pi^2 g_*(T)}\right)^{1/4}\rho_{R}(\varGamma_\phi,N_{re},n_{s}^k)^{1/4}~.
\eea
Combining this equation with eq.\ref{eqtre} we can establish one to one correspondence between $T_{re}$ and $\Gamma_{\phi}$ supplemented with the following condition for reheating 
\bea \label{reheating3}
H^2= \dot{N}_{re}= \rho_\phi(\Gamma_\phi,N_{re},n_{s}^k)+ \rho_{R}(\Gamma_\phi,N_{re},n_{s}^k))+\rho_{X}(\Gamma_\phi,N_{re},n_{s}^k)=\Gamma_{\phi}^2~~.
\eea

As we mentioned earlier that our another interest is to constraint dark matter phenomenology through the swampland conjecture. Therefore, while solving the Boltzmann equation we also need to consider the following condition on the dark matter abundance parametrized by $\Omega_X$, which is expressed in terms of radiation abundance $\Omega_R$ ($\Omega_R h^2=4.3\times10^{-5}$), as
\bea
\Omega_X h^2= \langle E_X\rangle \frac{X(T_F)~ T_F ~A_F}{R(T_F) ~T_{now} ~m_\phi} \Omega_R h^2~ = 0.12,
\eea
where $T_F$ is the temperature at a very late time when co-moving dark matter, as well as radiation density, became constant. 
The present value of dark matter abundance imposes a constraint on the dark matter parameter space $(M_X,\langle \sigma v \rangle)$ by the 
CMB anisotropies through the scalar spectral index $n_s^k$ considering swampland conjecture.

Therefore, inflationary dynamics, and considering the CMB temperature anisotropy, we will be able to constrain reheating as well as dark matter parameter space though the swampland conjecture. With all the ingredient discussed so far, and considering the condition of the refined swampland conjecture 
Eq.\ref{swamp1}, we can figure out the allowed region of ($c,c'$) with respect to inflation parameters  $(n_s^k,r_k)$. For our discussions, we will be considering $(c,c')$, as free parameters. The region of $n_s^k$ will be considered to be bounded by the PLANCK $2\sigma$ region in $(n_s^k,r)$ plane. With this consideration, a particular value of $c$, in general provides the maximum allowed value of $n_s^k$ and that in turn, imposes restriction not only on the maximum value of the reheating temperature ($T_{re}^{max}$) 
but also on the dark matter parameter space allowed by current dark matter abundance. Similarly, the maximum allowed value of reheating temperature, which is associated 
 with a particular value of the spectral index, should also impose constraints on c. We show the resulting constraints on $c$ and $c'$ for the maximum value of the scalar spectral index ($n_s^{max}$) 
 and the minimum value of the scalar spectral index ($n_s^{min}$). In the following 
 discussions, we consider various inflation model and discuss important results  
 of our analysis. In all cases, we consider two different effective equation 
 of state parameter for reheating, $\omega_{re}=0$ and $\omega_{re}=\frac{1}{6}$.
 \subsection{Chaotic inflation  \cite{Linde:1983gd}}
 
 \begin{figure}[t!] \label{chaotic-plot}
 	\begin{center}
 		\includegraphics[width=007.0cm,height=05.5cm]{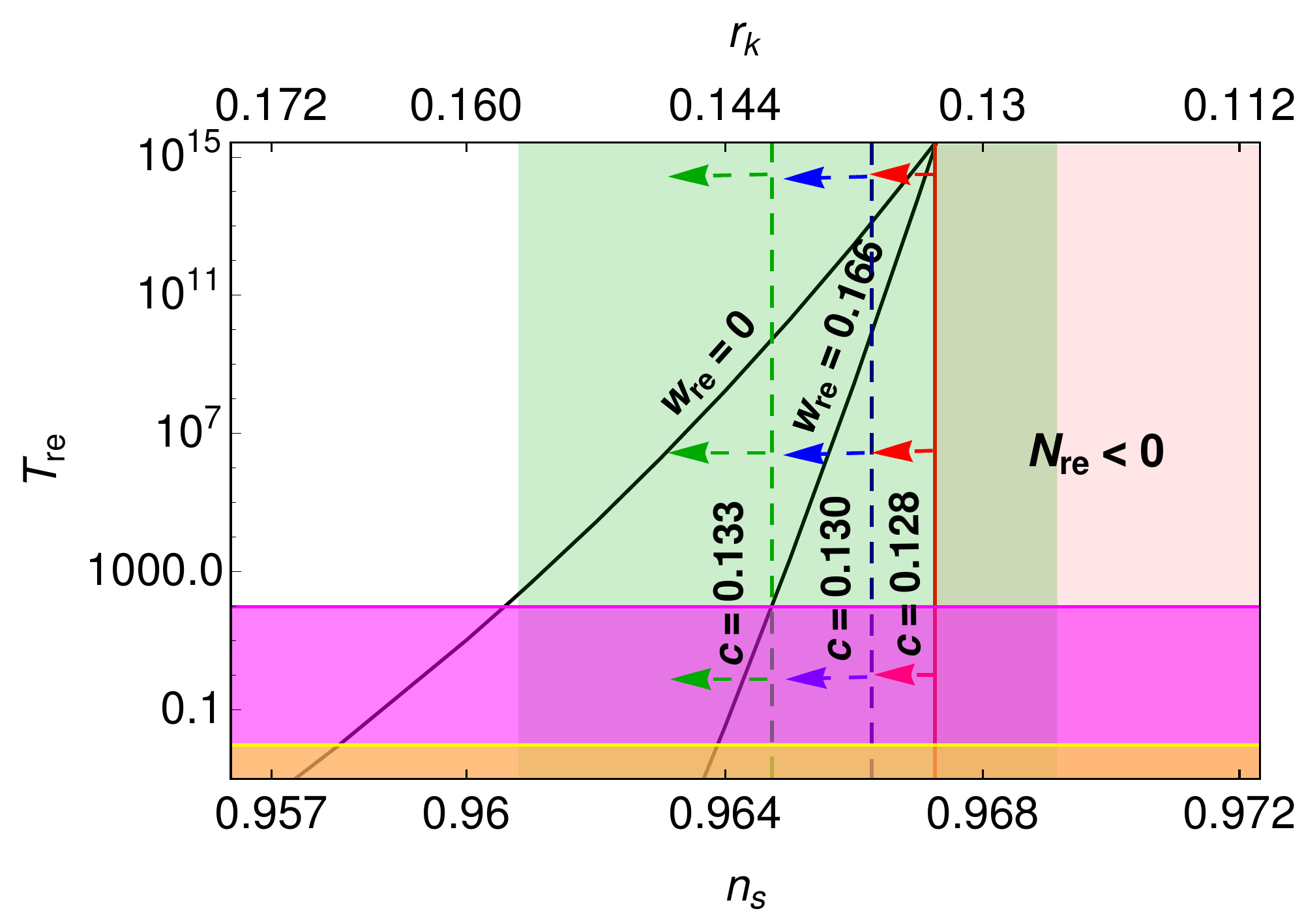}
 		\includegraphics[width=007.0cm,height=05.5cm]{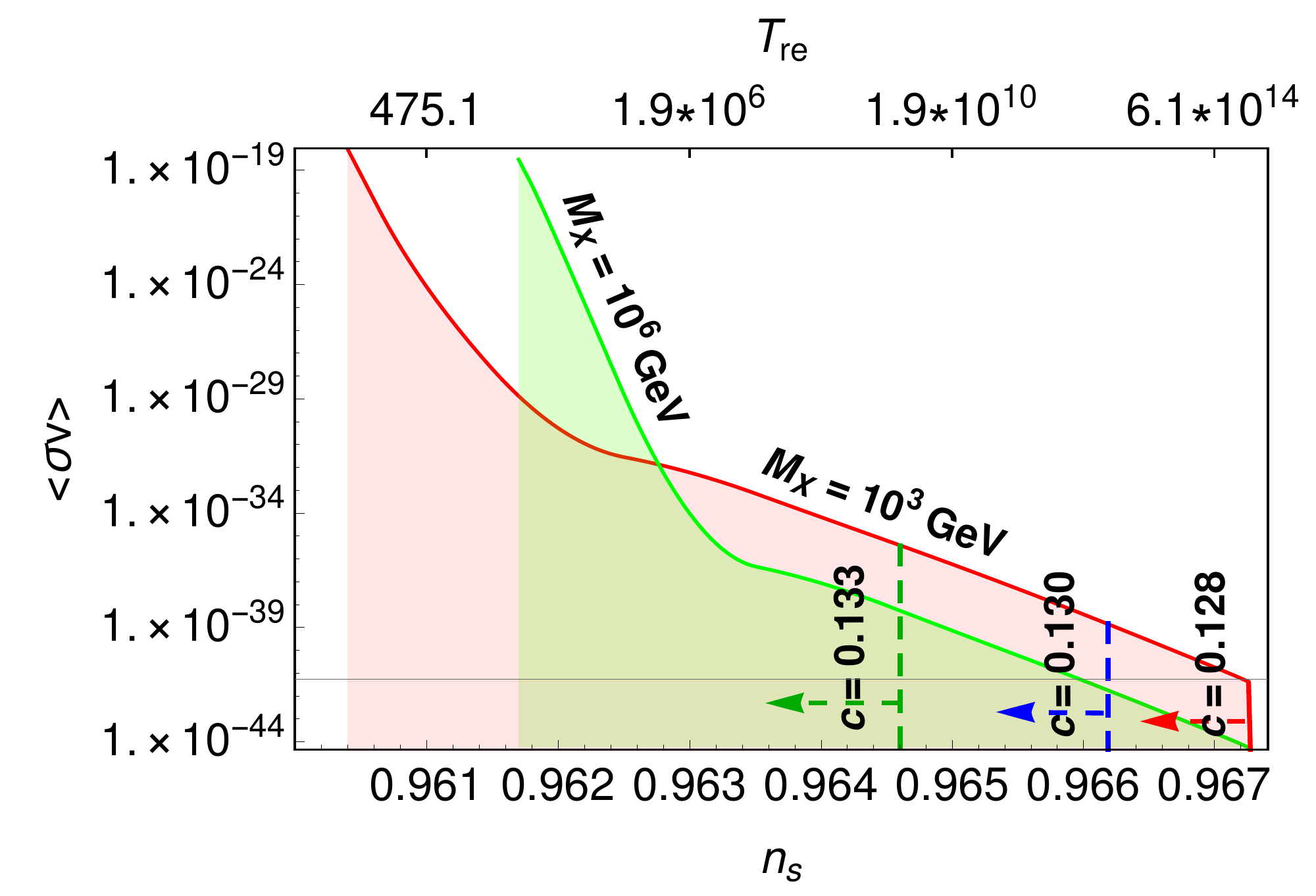}
 		\includegraphics[width=006.8cm,height=04.9cm]{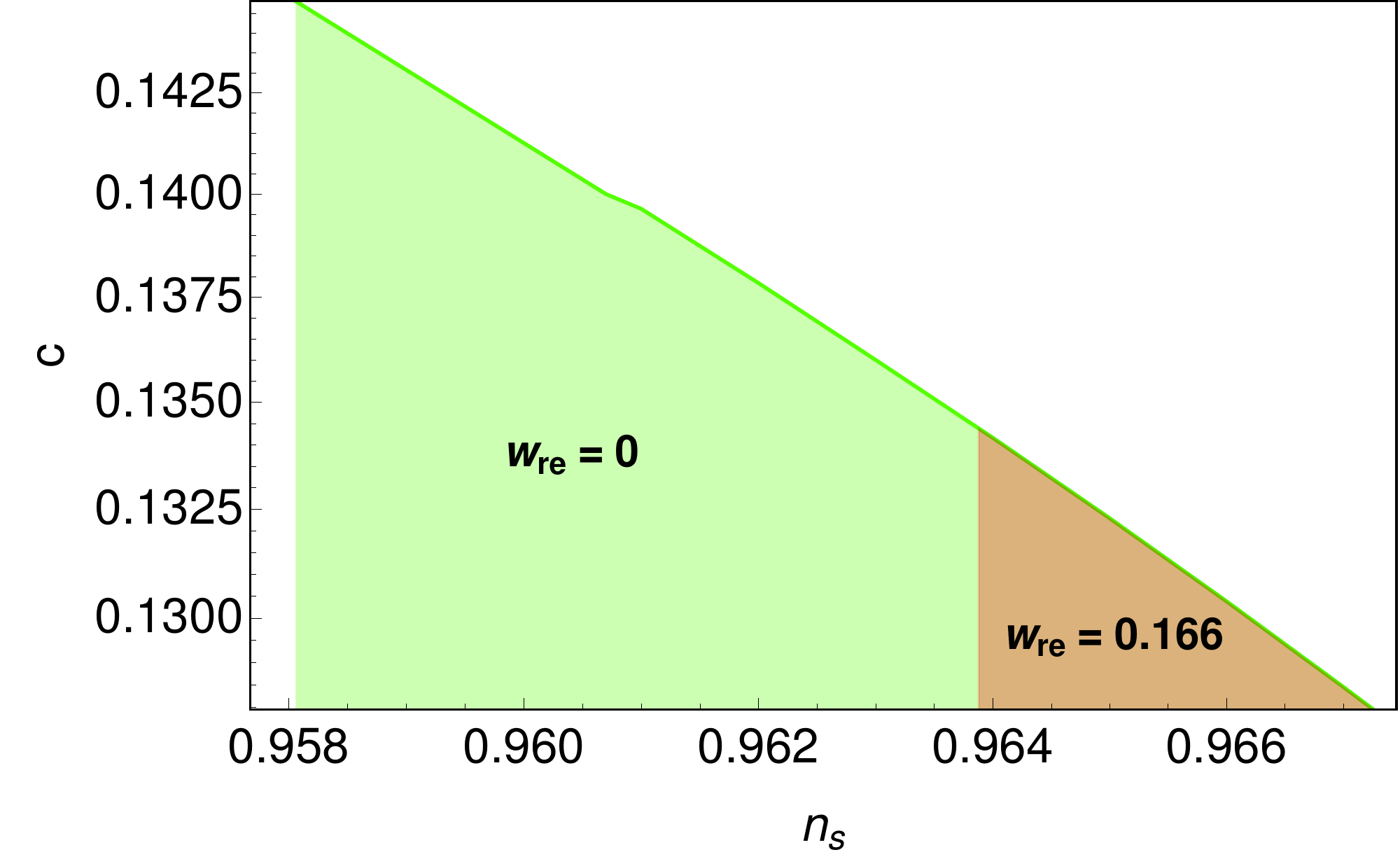}
 		\includegraphics[width=006.8cm,height=04.9cm]{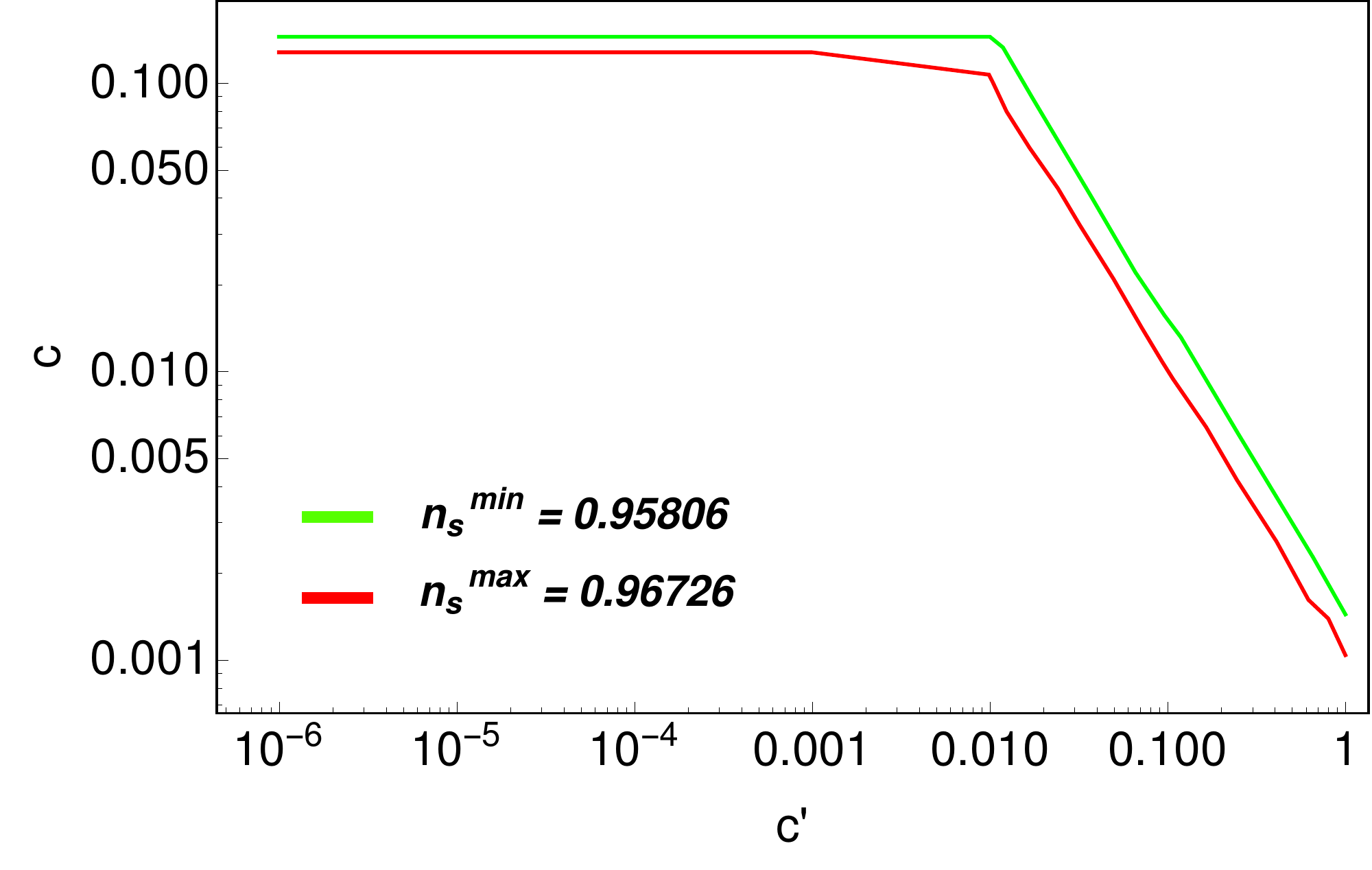}
 		\caption{\scriptsize We plot on the  upper left side variation of the reheating temperature ($T_{re}$) as a function of the spectral index($n_s$)
 		 for two effective equation-of-state, $\omega_{re}=0$ and $\omega_{re}=\frac{1}{6}$. A red solid line indicates
 		the maximum value of the spectral index ($n_s^{max}$) which is corresponds
 		 to c=0.128. The intersection point of the temperature curves and the red solid line indicates the maximum value 
 		 of the reheating temperature ($T_{re}^{max}$). Temperature above 
 		 the intersection point is unphysical as they correspond to 
 		 $N_{re}<0$ (light red region). The light green band indicates $1\sigma$ 
 		 range, $n_s=0.9649\pm0.0042$ (68 \% CL, Planck TT,TE,EE+lowE+lensing) from Planck \cite{Akrami:2018odb}
 		  and the dark pink region is 
 		  below the electroweak scale $T_{ew}\sim 100$ GeV. The dark yellow region below $10^{-2}$ GeV, 
 		  would ruin the predictions of big bang nucleosynthesis (BBN). Different  (shown in different colours dashed line) values of $c$ impose a
 		 further restriction on the maximum value of the reheating temperature. Upper right side, we have plotted the contour of 
 		 $\Omega_X h^2=0.12$ in $n_s-\langle\sigma v \rangle$ plane for fixed dark matter masses.
 		  The shaded region below the contour line is the allowed parameter space, that is further constraint by different values of c (shown in different colours).
 		  On the lower left side, we plot allowed values of $c$ as a function of 
 		 $n_s$ within maximum and minimum values of the reheating temperature. The region under 
 		 the green solid line is for $\omega_{re}=0$ and dark red region for $\omega_{re}=\frac{1}{6}$. In the lower right side, we show the resulting constraints for $c$ and 
 		$c'$ for maximum and minimum values of the spectral index with considering effective equation of state $\omega_{re}=0$. Minimum and maximum values of the spectral index correspond to $T_{re}\sim 10^{-2}$ GeV and $N_{re}\sim 0$ respectively.
 		 All four plots 
 		are for the chaotic inflation model with $n=2$.}
 		\label{chaotic}
 	\end{center}
 \end{figure}
 For usual chaotic inflation model potential looks like
 \bea \label{chao}
V(\phi)= \frac 1 2 m^{4-n} \phi^n~~,
\eea
where $n = 2,4,6 \dots$. If we consider the absolute value of the field, $n=3,5,\dots$ can also be included. 
We consider only $n=2$ for our numerical purpose. 
As mentioned earlier, the first condition of 
the refined swampland conjecture, $M_p {V'}/{V}$, transforms into the following inequality,
\bea \label{swamp3}
M_p \frac{V'}{V}=M_p \frac{2}{\phi_k}\geq c~~,
\eea
where, inflaton field value for a particular scale of interest $k$, can be written as  
\bea \label{phik}
\phi_k^2= \frac{1}{1-n_s^k}\left(3 M_p^2 n^2-2n(n-1)M_p^2\right)~~.
\eea
After combining equations (\ref{swamp3}) and (\ref{phik}), we see 
that a particular value of $c$ gives an upper bound on the value of scalar spectral index $n_s^k$. This constraint sets a maximum possible value of the reheating temperature ($T_{re}^{max}$). Because of implicit relation, the above constraint also sets a minimum possible value of dark matter scattering cross-section $\langle\sigma v\rangle$ for a given dark matter mass. Similarly 
from the second condition of the conjecture, we find the following further constraint on the inflaton field value, 
\bea \label{swamp4}
M_p^2 \frac{V''}{V}=\frac{2}{\phi_k^2} M_p^2\leq -c'~~.
\eea
Combing the equations (\ref{swamp3}), (\ref{phik}) and (\ref{swamp4}),  
we are able to find constraints on $c$, $c'$. For $n=2$, combine 
equation can be written as
\bea \label{swampc}
c^2 c'^2\leq \frac{1}{32}(1-n_s^k)^3~~.
\eea
By using the above constraint relation along with any one of the swampland conjecture say equation (\ref{swamp3}), we have the allowed region of the $(c,c')$ space as shown in Fig.\ref{chaotic}. 
The maximum value of the spectral index for this $n=2$ model is  $n_s^{max}\simeq 0.96726$, and it remains same for different effective reheating equation of state parameter $\omega_{re}$ as can be inferred from the first figure of fig.\ref{chaotic}. Most importantly associated with this maximum possible temperature we have  a unique value of $c =c^{{tmax}}\simeq 0.128$, which does not depend upon the value of effective equation of state. Most importantly from the predictions of the BBN, there exist minimum values of the spectral index $n_s^{min} \simeq (0.95806,~0.96388)$ for different values $\omega_{re}=(0,\frac{1}{6})$, which provides us maximum possible value of $c^{max} = (0.1448,0.1344)$.
{\em Most important result of our present analysis} is that for each  value of $c$ between $(c^{max} > c > c^{tmax})$, there exists a maximum allowed value of reheating temperature $T^{max}_{re}(c)$ for a given equation of state. For example $c=0.130$ corresponds to the maximum 
 allowed values of the reheating temperature to be $T^{max}_{re}(0.13) =(8.4\times10^{12}, 3.7\times10^{9})$ GeV for $\omega_{re}=(0,\frac{1}{6})$ respectively. 
In the third plot of fig.\ref{chaotic} we have shown the allowed region of the swampland parameter $c$ in terms of scalar spectral index $n_s$. Here we can see that allowed region of $c$ start decreasing with increasing effective equation of state $\omega_{re}$.\\
 In the fourth plot of fig.\ref{chaotic}, we have plotted annihilation cross section as a function of $n_s$ for different dark matter masses within the maximum and minimum values of the reheating temperature. 
 For a given value of $c$, one can precisely predict the values of the annihilation cross-section once dark matter mass is fixed. 
 As an example given the value $c=0.130$, the current relic abundance fixes the dark matter annihilation cross-section within $8.8\times10^{-19} ~GeV^{-2}> \langle\sigma v \rangle> 1.5\times10^{-39}~ GeV^{-2}$ for $M_X=10^3 \mbox{GeV}$ and 
 $2.3\times10^{-19} ~GeV^{-2}> \langle\sigma v \rangle> 1.8\times10^{-42}~ GeV^{-2}$ for $M_X=10^6$ $GeV$. Importantly considering the PLANCK observation, if we increase the value of $c$ towards $c^{max}$, this range of dark matter annihilation cross-section will be further narrowed down.
\subsection{Natural inflation \cite{axion}}
 The inflationary potential in this model is given by
\bea \label{naturalinflation}
V(\phi)=  \Lambda^4 \left[1 - \cos\left(\frac{\phi}{f}\right) \right] .
\eea
where, $\Lambda$ is the height of the potential setting the inflationary
energy scale, and $f$ is the width of the potential, known as the axion decay constant. 
The CMB normalization fixes the overall scale of the inflation $\Lambda$. 
Therefore, by tuning the value of the axion decay constant $f$ we can fit this
 model with observation. We have chosen two sample values of axion decay constant 
 $f=(10 M_p,50 M_p)$ and those values of decay constant are marginally consistent
  with Planck data. The maximum reheating temperature $T^{max}_{re}\sim 10^{15}$ GeV for different values of $f$ and $\omega_{re}$, arising due to instant reheating ($N_{re}\sim0$) 
  sets the maximum possible value of scalar spectral index $n_s^{max}\simeq (0.96617,~0.96726)$ for $f=(10M_p,50M_p)$ respectively. 
  However as already discussed for the chaotic inflation, the maximum possible value of $c=c^{max}$ arises from the minimum possible values of spectral index which again corresponds to minimum value of the reheating temperature. The Minimum values of the spectral index  $n_s^{min} \simeq (0.95717,~0.9629)$ for $(\omega_{re},f)=((0,10 M_p), (\frac{1}{6},10 M_p))$ and 
  $n_s^{min} \simeq (0.9581,~0.9639)$ for $(\omega_{re},f)=((0,50 M_p) , (\frac{1}{6},50 M_p))$.
 With these ingredients the constraints on $c$ will be obtained from swampland conjecture Eq.\ref{swamp1} and Eq.\ref{swamp4} which transforms into the following inequalities for the axion inflaton,
    \bea \label{swampnatural1}
    M_{p} \frac{V'}{V}&=&\frac{M_p}{f}~ \cot\left(\frac{\phi}{2f}\right)=\frac{M_p}{f}\sqrt{\frac{f^2}{M_p^2}(1-n_s)}~\geq c~~,\\
M_p^2\frac{V''}{V}&=&\left(\frac{f^2(1-n_s)-3M_p^2}{4f^2}\right)\leq~-c'~~.
\eea
\begin{figure}[t!]
 	\begin{center}
 		\includegraphics[width=005.41cm,height=04.1cm]{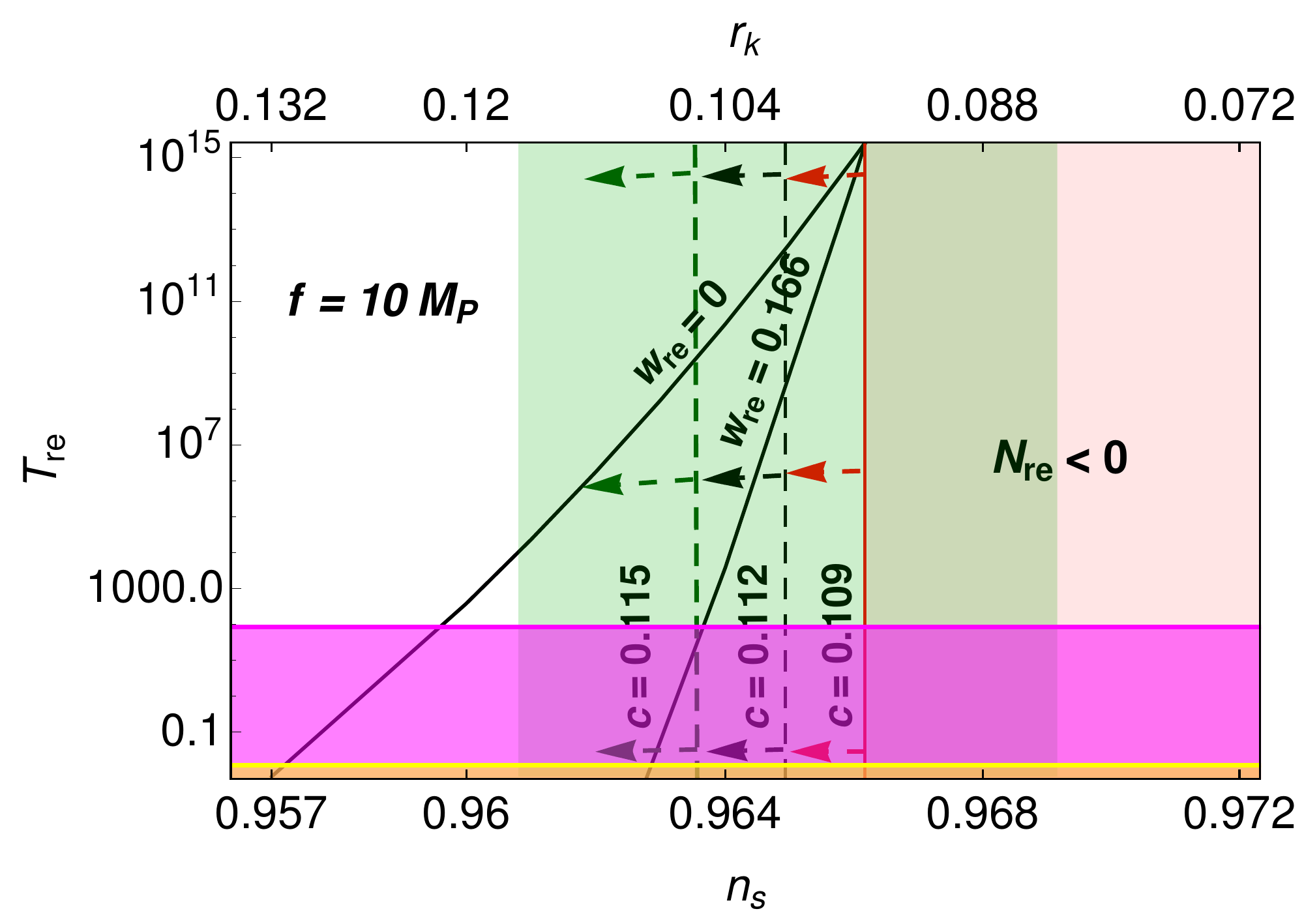}
 		\includegraphics[width=005.41cm,height=04.1cm]{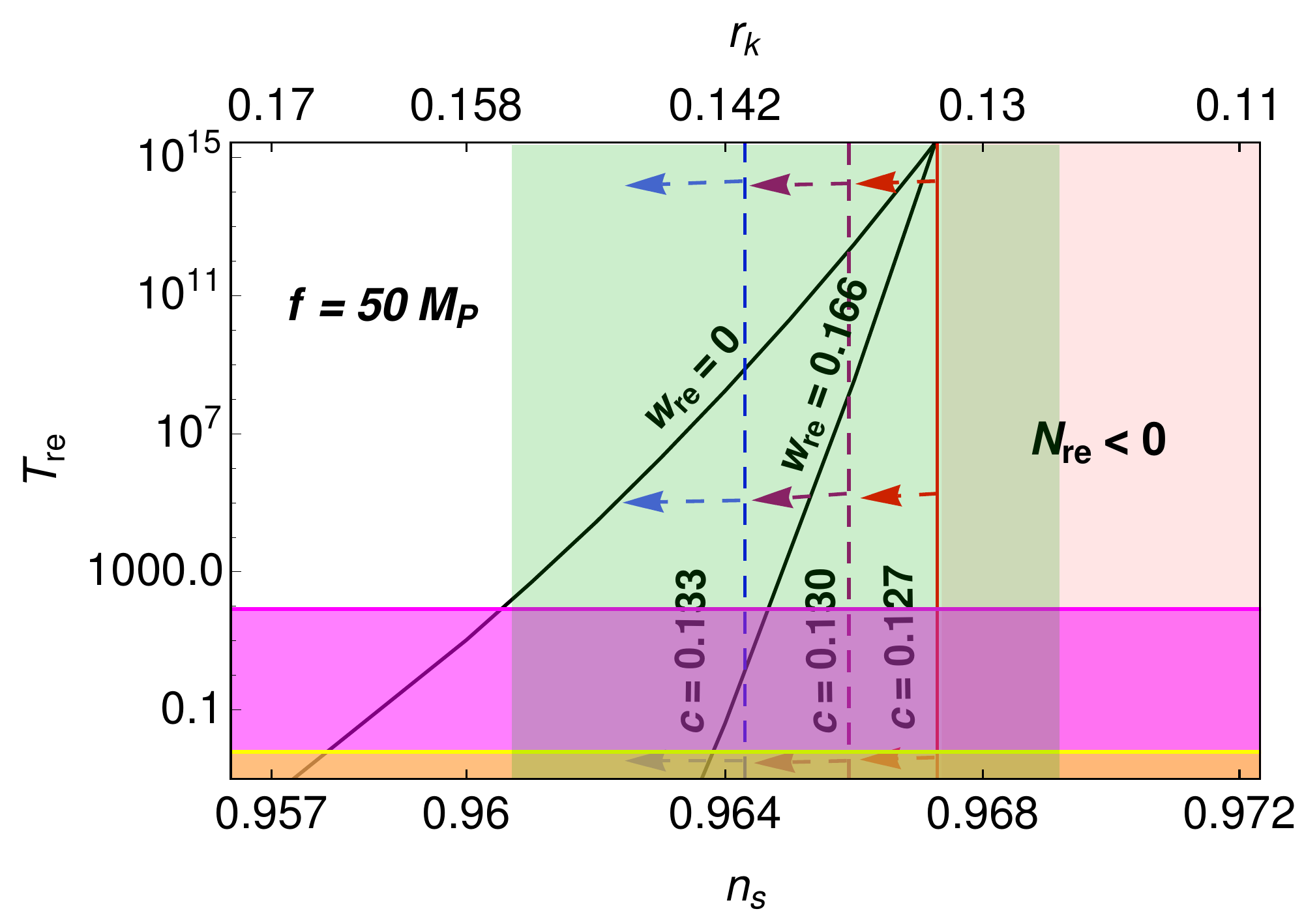}
 		\includegraphics[width=005.41cm,height=04.1cm]{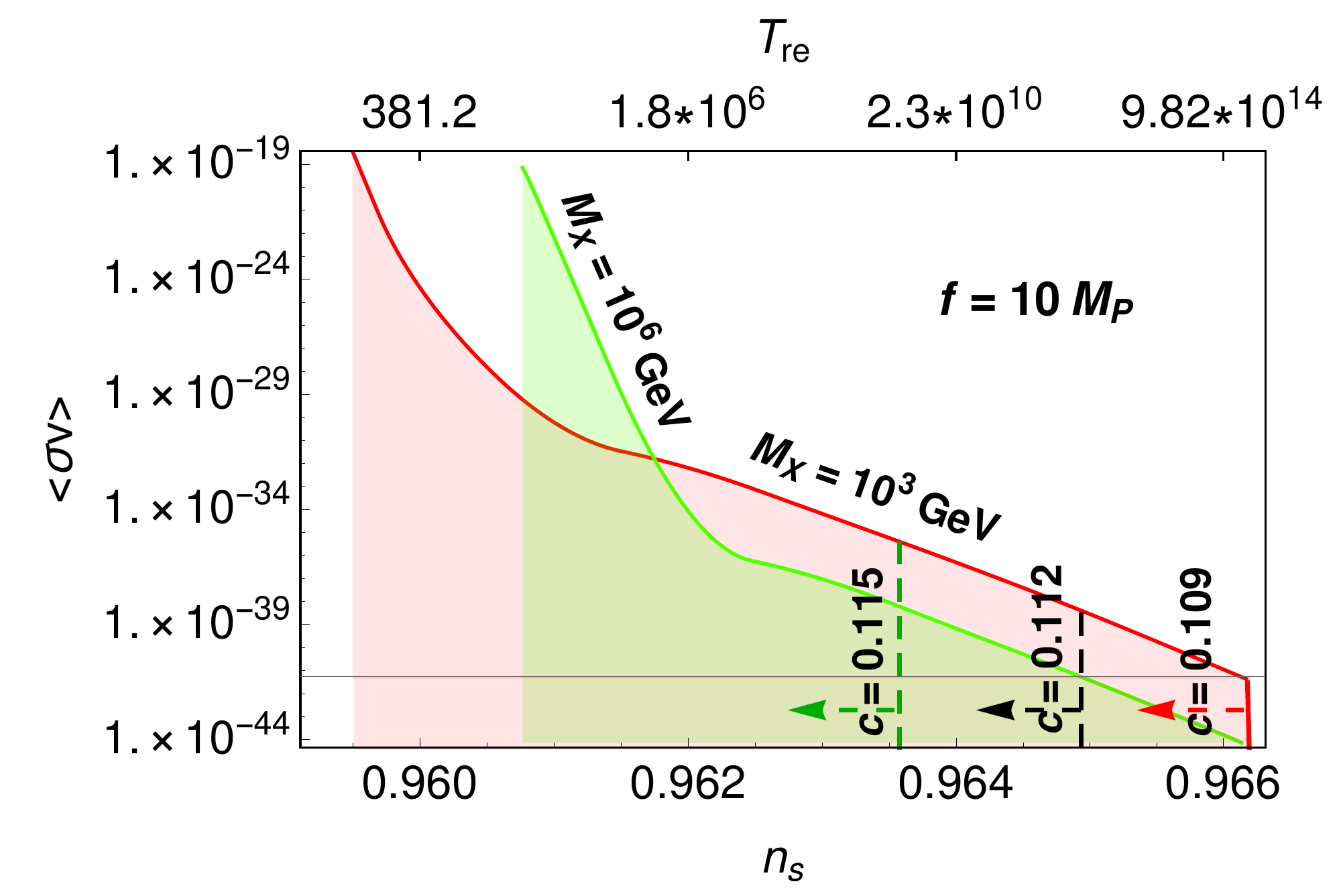}
 		\includegraphics[width=005.41cm,height=04.1cm]{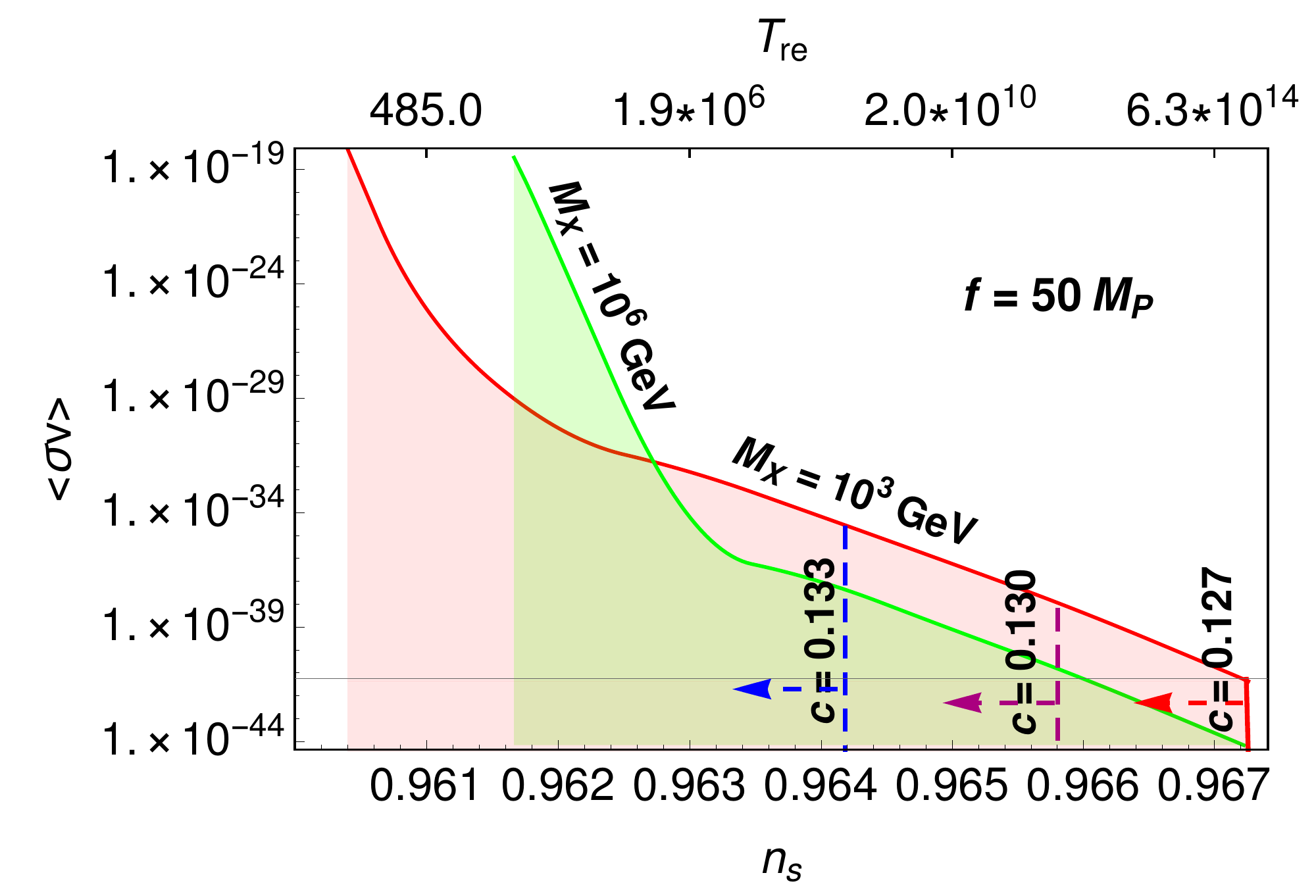}
 \includegraphics[width=005.4cm,height=03.6cm]{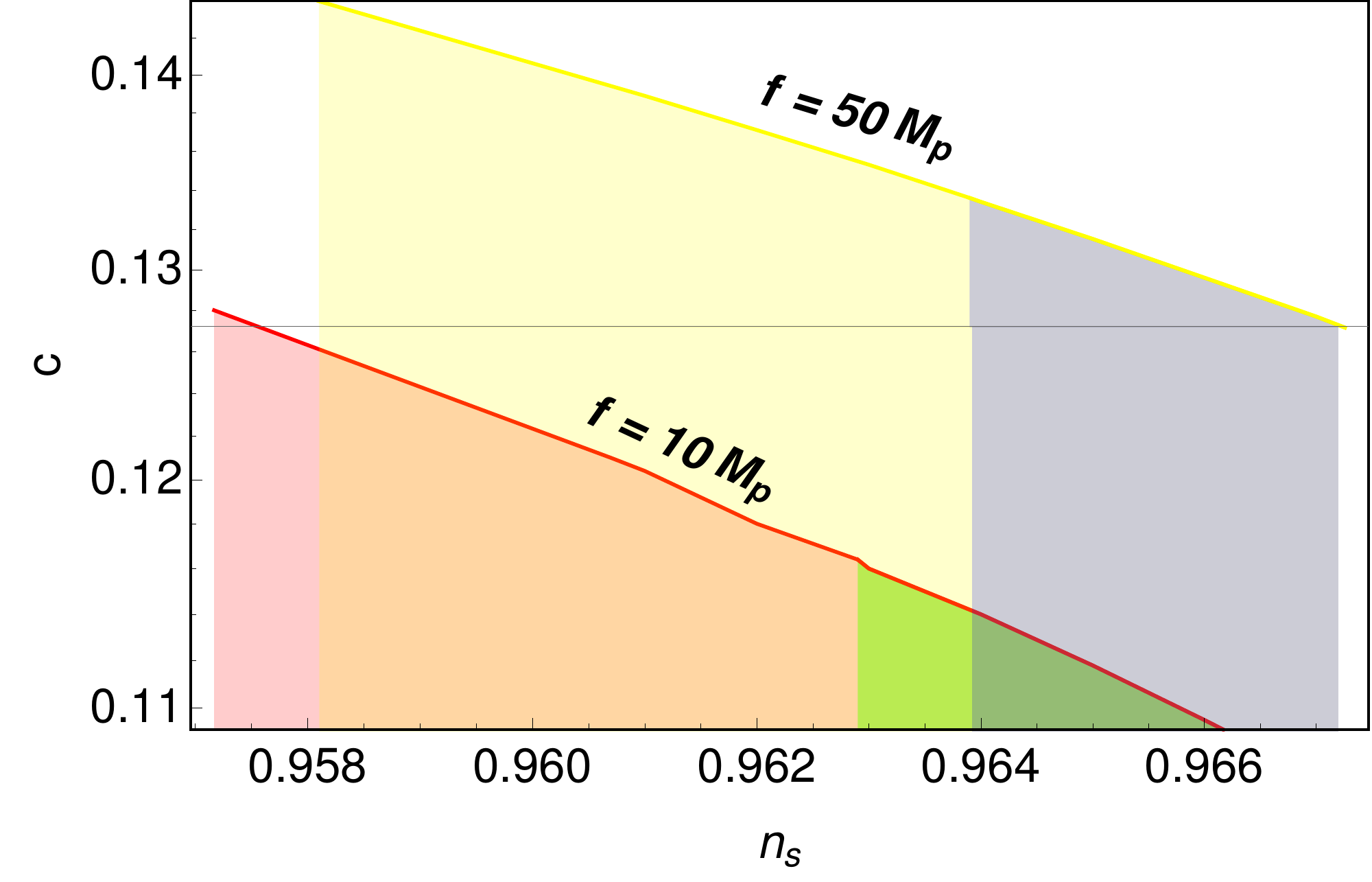}
 		\includegraphics[width=005.4cm,height=03.6cm]{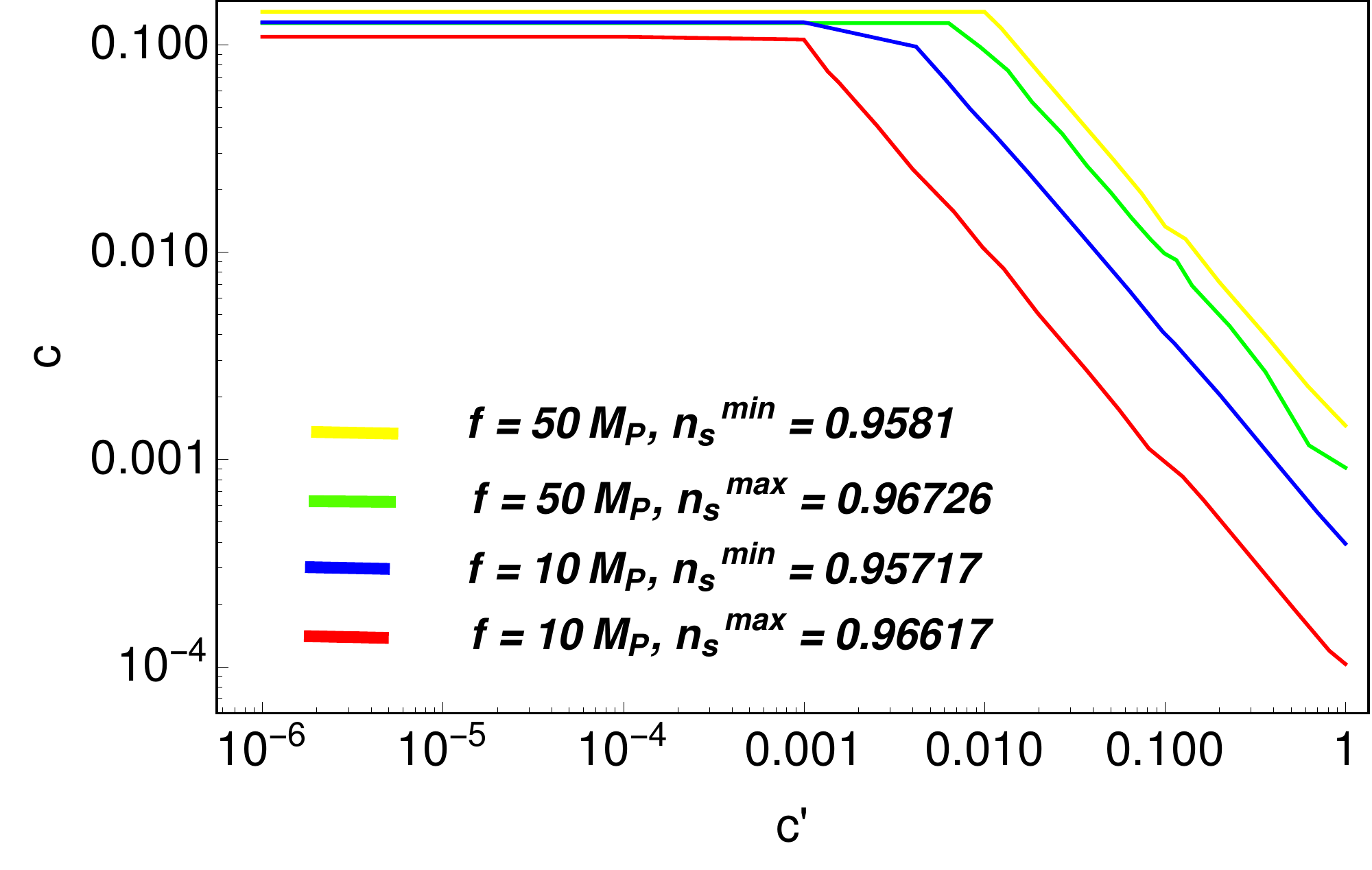}
 		\caption{\scriptsize All plots are same as in the previous fig.\ref{chaotic}. The only difference is that here we have plotted for natural inflation model 
 		for $f=(10 M_p, 50 M_p)$. In the lower middle plot, the region under 
 		 the yellow solid line is for ($\omega_{re}, f)=(0, 50 M_p)$ and dark blue region is for ($\omega_{re}, f)=(\frac{1}{6}, 50 M_p)$. In the 
 		 same plot, the region under 
 		 the red solid line is for ($\omega_{re}, f)=(0, 10 M_p)$ and dark green region is for ($\omega_{re}, f)=(\frac{1}{6}, 10 M_p)$.}
 		\label{NATURAL}
 	\end{center}
 \end{figure}
   One important fact to notice is that, at the point of instantaneous reheating ($N_{re}\sim0$), the associated value of $c$ ($c^{tmax}$) is independent of reheating equation of state but dependent on the axion decay constant $f$, such as    $c^{tmax}=(0.109,~0.127)$ for $f=(10,50)M_p$ for both values of equation of state. As was the case for chaotic inflation, eq.\ref{swampnatural1}  which constrains the value of spectral index as $n_s<1-(2c^2+\frac{M_p^2}{f^2})$, will directly restrict the possible value of the reheating temperature given a value of $c> c^{tmax}$. For 
  a fixed values of $c$ (greater than $c^{tmax}$) and $f$, maximum allowed values of reheating temperature are different for different equation of state $\omega_{re}$. Considering $(f,c)=(50M_p,0.130)$, the maximum allowed values of reheating temperature turns out to be  $T_{re}^{max} =(1.1\times10^{12}, 3.1\times 10^7)$ GeV for $\omega_{re}=(0,{1}/{6})$ respectively. 
  Similarly maximum and minimum  values of reheating temperature from observation impose a restriction on the allowed values of $c$, which we have shown in the $(c~vs~n_s)$ plot in Fig.\ref{NATURAL}. 
 Combining two equations (\ref{swampnatural1}) one finds
 \bea
 c^2 c'^2~\leq \frac{1}{32} \left(\frac{f^2(1-n_s)-3 M_p^2}{f^3}\right)^2 (f^2(1-n_s)- M_p^2)~~.
 \eea
 The resulting constraints in $(c, c')$ space has been plotted in the last plot of the fig.\ref{NATURAL}, where the upper limit on $c$ has been derived from the PLANCK constraints.\\
Now we turn to understand the dark matter parameter space, specifically how the parameter $c$ constrains the dark matter annihilation cross-section, 
  $\langle \sigma v\rangle$, for different scalar spectral index or reheating temperature. In the same way as we have seen in the chaotic inflation 
  model, for a fixed values of $(c,f)$, for example $c=0.130$ and $f=50 M_p$, the annihilation cross-section will be constrained within $7.9 \times 10^{-19}~ \mbox{GeV}^{-2}> \langle\sigma v \rangle> 1.4 \times 10^{-38} ~\mbox{GeV}^{-2}$, considering  $M_X=10^3$ GeV and 
 $2\times10^{-19} ~\mbox{GeV}^{-2}  >  \langle\sigma v \rangle > 1.8\times10^{-41}~ \mbox{GeV}^{-2}$ for $M_X=10^6 ~\mbox{GeV}$.  
 As has already been observed for chaotic inflation, the lower limit of annihilation cross-section ($\langle\sigma v \rangle ^{lower}$) will be higher for higher values of $c$, and it becomes narrowed  like for $c=0.133, 
 \langle\sigma v \rangle ^{lower}= (2.16\times10^{-35},4.4\times 10^{-38}) $ $ \mbox{GeV}^{-2}$ for $M_X=(10^3,10^6)$ $\mbox{GeV}$.

   \subsection{\bf $\alpha-$attractor model \cite{alpha}}
 
  In this section, we will consider a class of models which unifies a large 
  number of inflationary models parameterized by a parameter $\alpha$, 
  first proposed in \cite{alpha}. Conformal property of this class of models leads to a
   universal prediction for the inflationary observables. In the canonical form, the so-called $\alpha$-attractor potential is expressed as
    \begin{equation} \label{a}
V(\phi) = \Lambda^4 \left[  1 - e^{ -     \sqrt{\frac{2}{3\alpha} }    \frac{\phi}{M_p}     } \right]^{2n}.
\end{equation}
 This model is known as E-model. The mass scale $\Lambda$ can be fixed from the 
 CMB power spectrum. The parameter $\alpha$ determines the shape of the canonically normalized inflaton potential near its minimum. This model includes 
 the starobinsky model for $n=1$ and $\alpha=1$. 
 In our present analysis we consider two cases with $n=1$ and $\alpha=(1,100)$ for comparison. For this choice of parameters both maximum and minimum values of the spectral index corresponding to  maximum and minimum values of reheating temperature respectively lie 
 within the $2\sigma$ range of PLANCK. The maximum value of the spectral index turns out to be $n_s^{max}\simeq (0.96717,0.9702)$ for $\alpha=1$ 
 and $100$ respectively. Once again the value of the $n_s^{max}$ is the same for a different equation of state $\omega_{re}$ once the parameter $\alpha$ is fixed. From the predictions of big bang nucleosynthesis (BBN), 
  the minimum values of  spectral index $n_s^{min}$ is different for a different $\omega_{re}$ and $\alpha$, such as $n_s^{min}\simeq (0.9579,~0.96375)$ for $(\omega_{re},\alpha)=((0,1),(\frac{1}{6},1))$ and 
  $n_s^{min}\simeq (0.9616,~0.96704)$ for $(\omega_{re},\alpha)=((0,100),(\frac{1}{6},100))$.
  \begin{figure}[t!]
 	\begin{center}
 		\includegraphics[width=005.41cm,height=04.1cm]{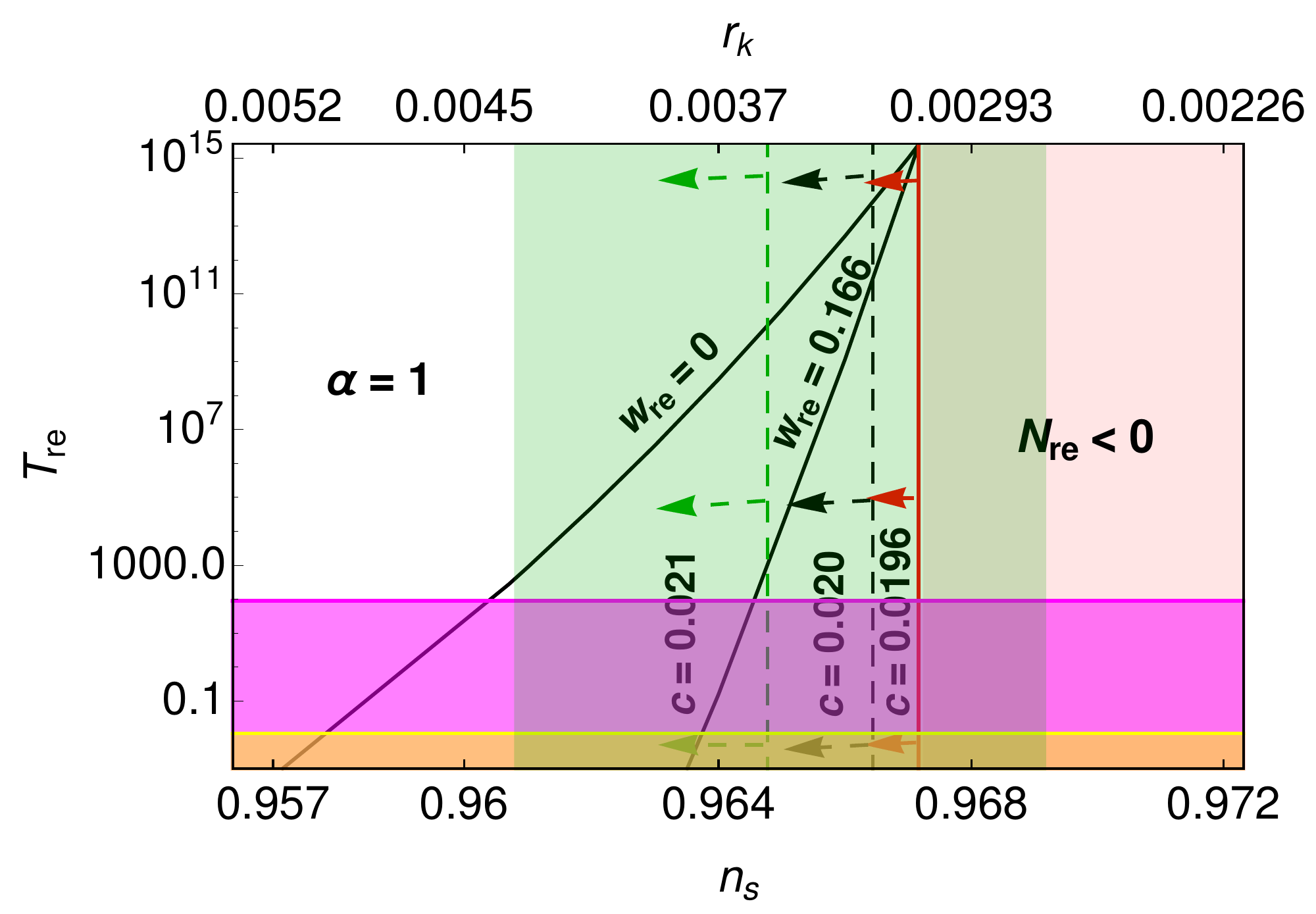}
 		\includegraphics[width=005.41cm,height=04.1cm]{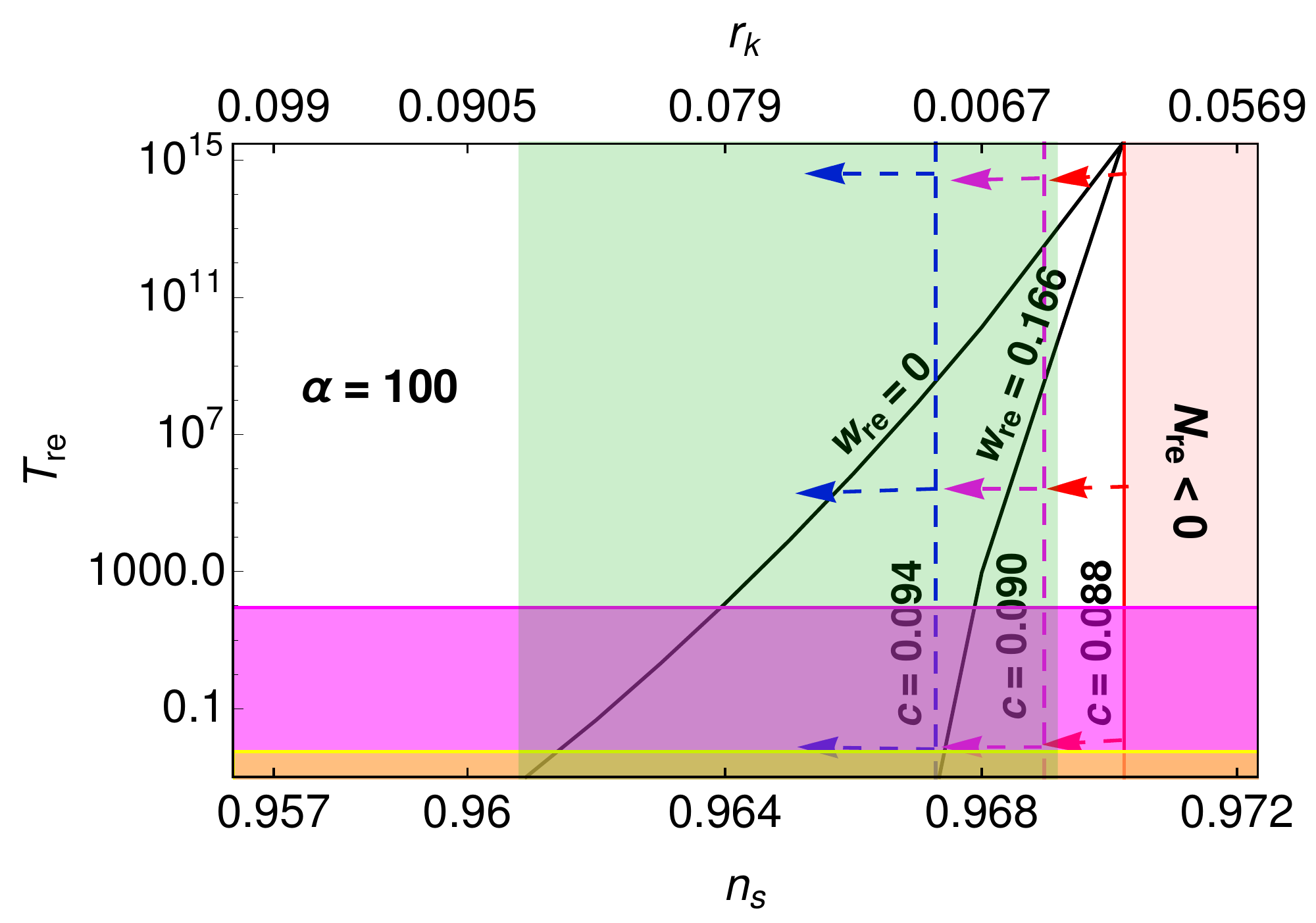}
 		\includegraphics[width=005.41cm,height=04.1cm]{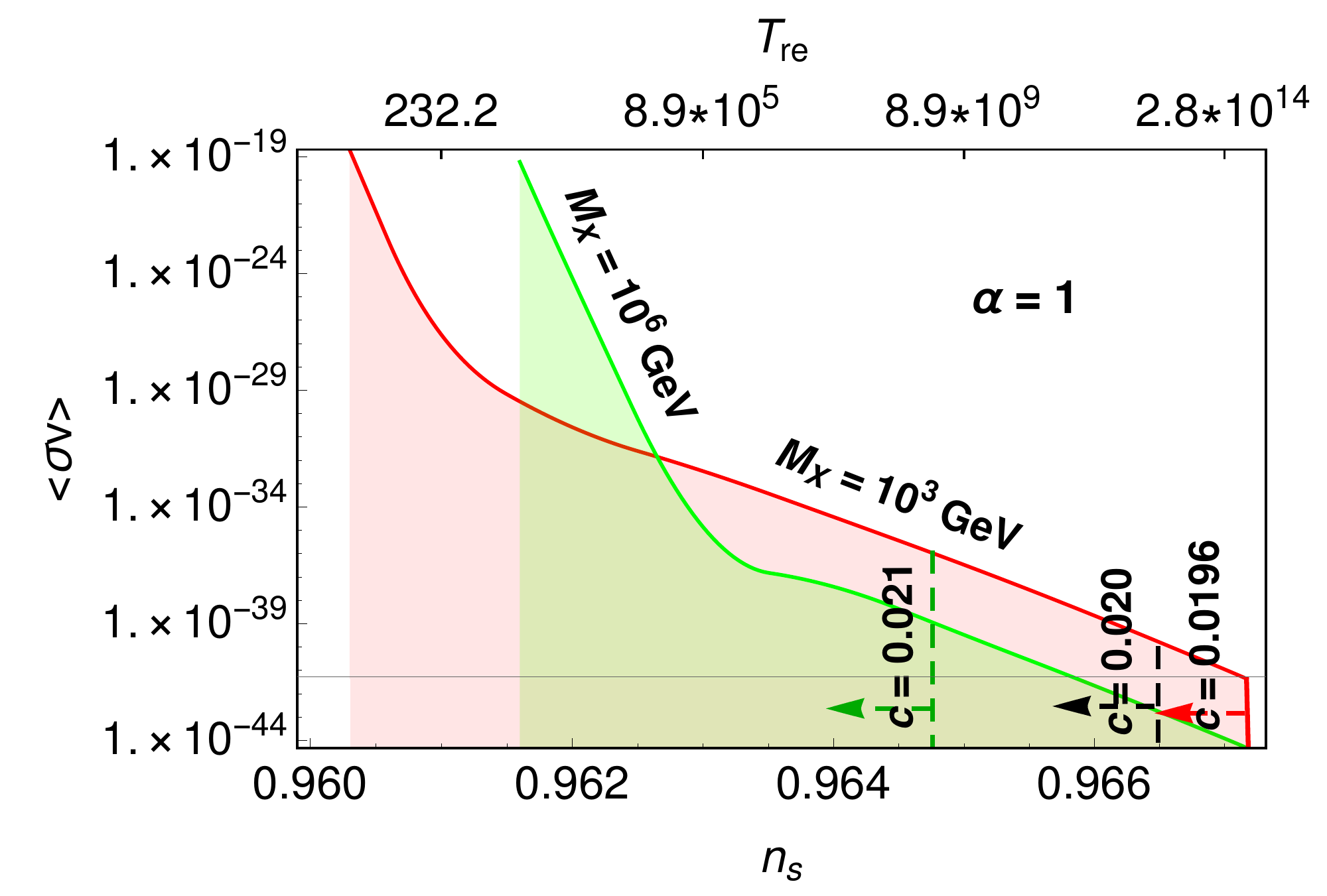}
 		\includegraphics[width=005.41cm,height=04.1cm]{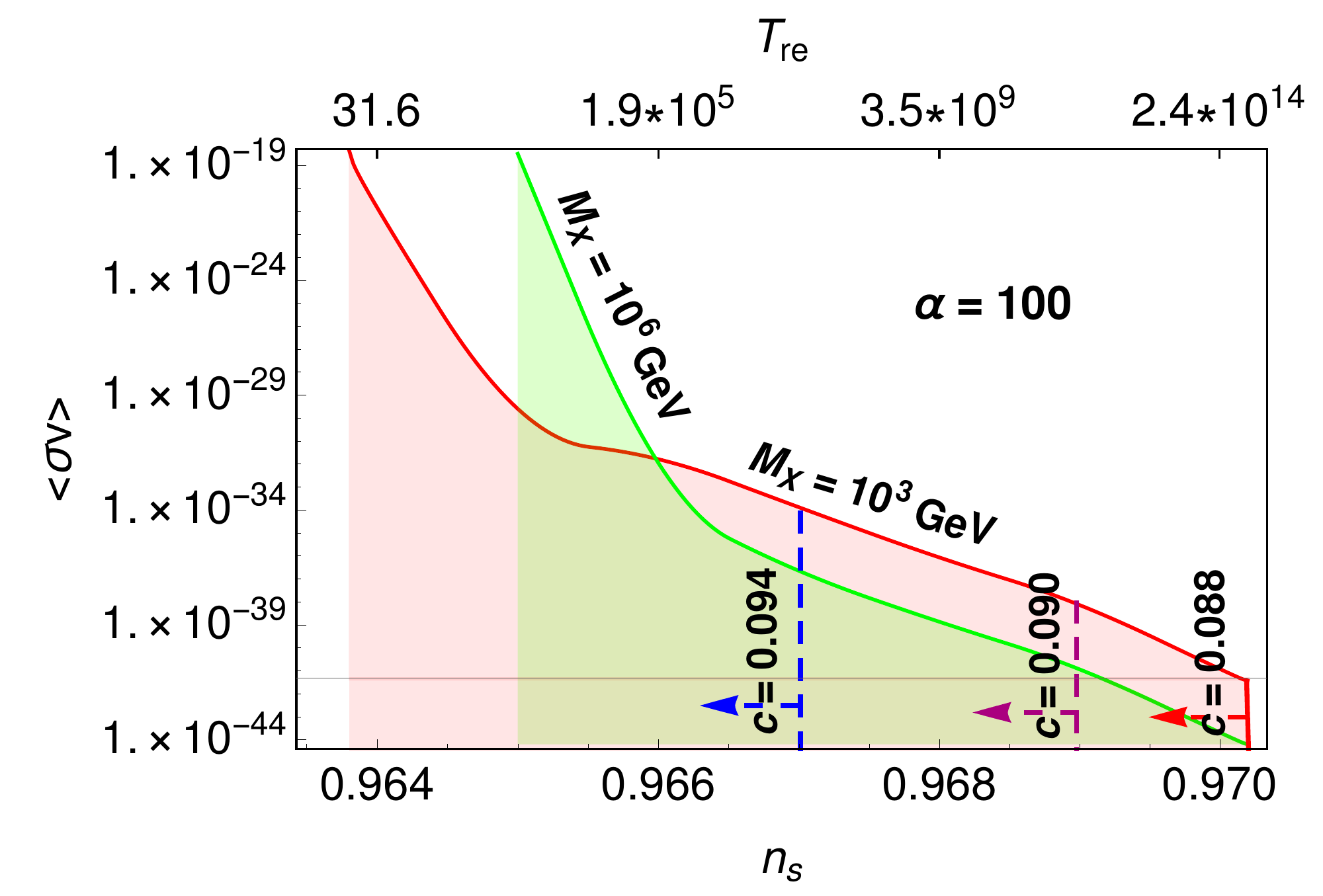}
 		\includegraphics[width=005.4cm,height=03.6cm]{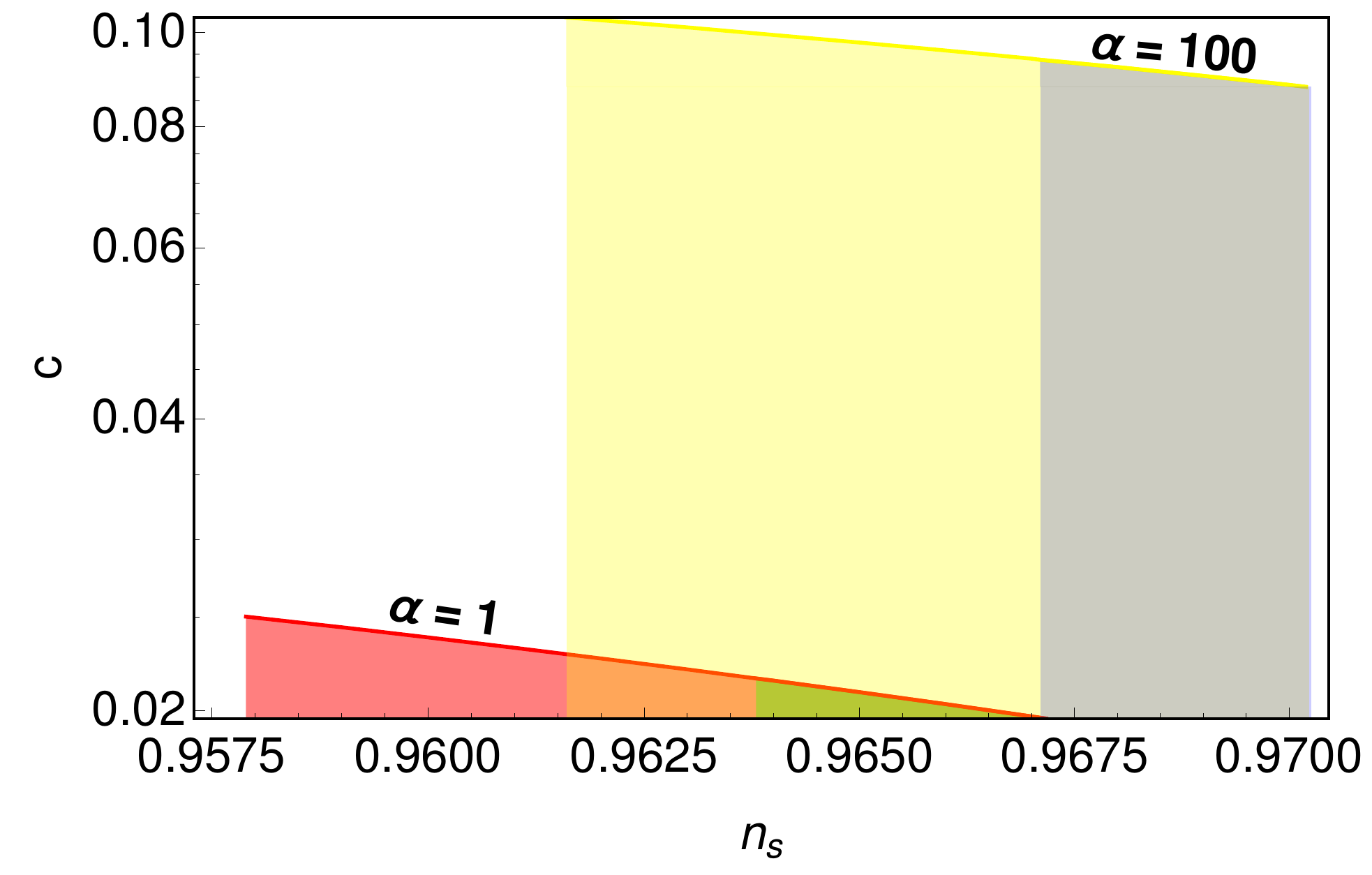}
 		\includegraphics[width=005.4cm,height=03.6cm]{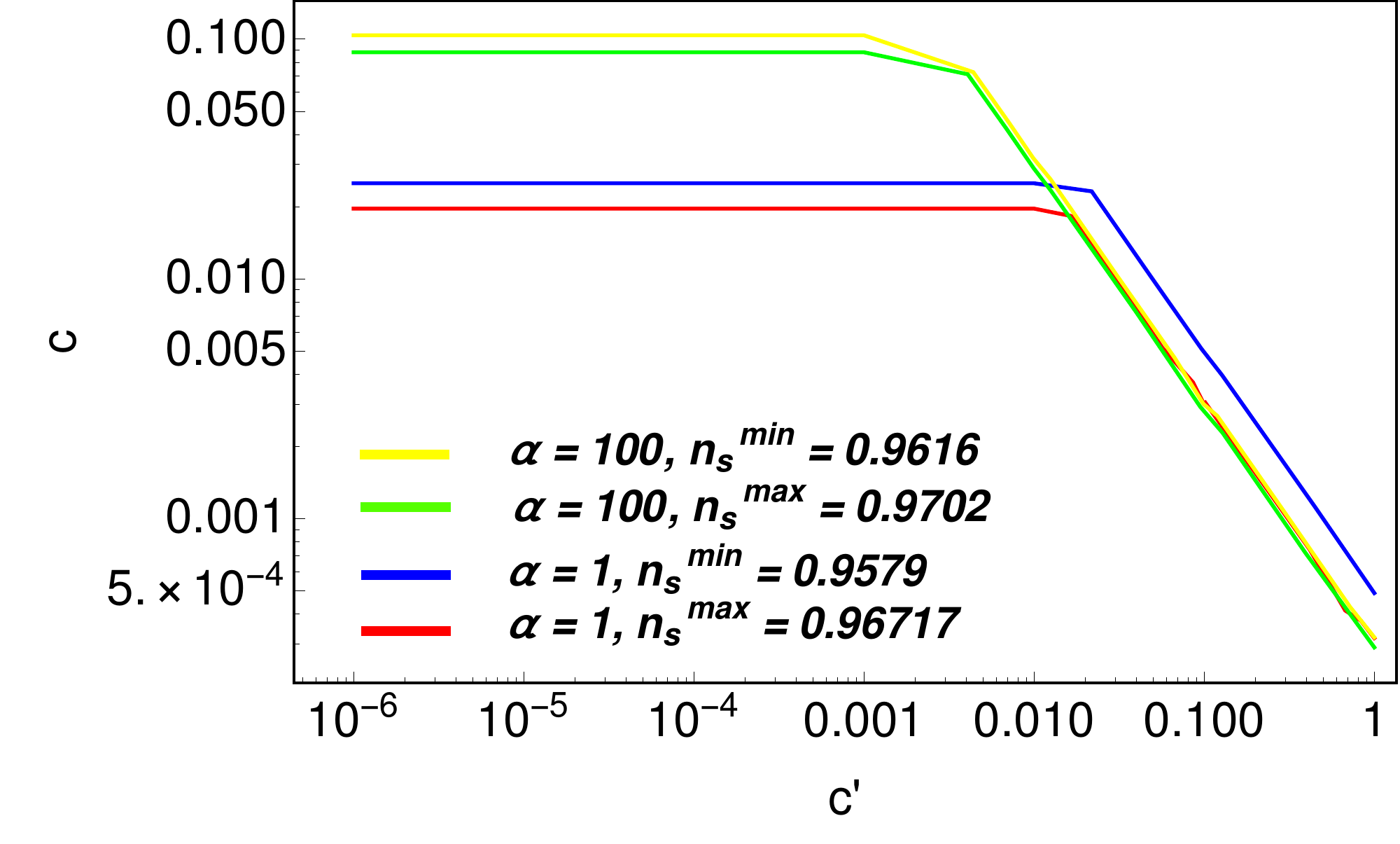}
 		\caption{\scriptsize All plots are same as in the previous fig.\ref{chaotic}. The main inequality is that here we have plotted for $\alpha$-attractor inflation model 
 		for $\alpha=(1, 100)$. In the lower middle plot, the region under 
 		 the yellow solid line is for ($\omega_{re}, \alpha)=(0, 100)$ and dark blue region is for ($\omega_{re}, \alpha)=(\frac{1}{6}, 100)$. In the 
 		 same plot, the region under 
 		 the red solid line is for ($\omega_{re}, \alpha)=(0, 1)$ and dark green region is for ($\omega_{re}, \alpha)=(\frac{1}{6}, 1)$.}
 		\label{compare}
 	\end{center}
 \end{figure}
 
 From the conditions of the refined swampland conjecture (\ref{swamp1}), the scalar spectral index ($n_s$) satisfy either
 \bea \label{alpha1}
 M_p~ \frac{V'}{V}=\frac{2n~ \sqrt{\frac{2}{3\alpha}}~e^{-\sqrt{\frac{2}{3\alpha}}\frac{\phi}{M_p}}}{1-e^{-\sqrt{\frac{2}{3\alpha}}\frac{\phi}{M_p}}}~\geq~c~~,
 \eea
 or
 \bea \label{alpha1'}
 M_p^2~ \frac{V''}{V}=\frac{4n}{3\alpha}~ e^{-\sqrt{\frac{2}{3\alpha}}\frac{\phi}{M_p}}\left(\frac{2n~ e^{-\sqrt{\frac{2}{3\alpha}}\frac{\phi}{M_p}}-1}{\left(1-e^{-\sqrt{\frac{2}{3\alpha}}\frac{\phi}{M_p}}\right)^2}\right)~\leq~-c'~~,
\eea
where $\phi$ can be written in terms of spectral index ($n_s$) as \cite{Drewes:2017fmn}
\bea \label{alpha2}
\phi=\sqrt{\frac{3\alpha}{2}}~M_p~ln\left(1+\frac{4n+\sqrt{16~n^2+24~\alpha ~n~(1+n)~(1-n_s)}}{3~\alpha~(1-n_s)}\right)~.
\eea
  As in the natural inflation model, here also we have $c^{tmax}$ corresponding to instantaneous reheating, which are independent of equation of state but dependent on  $\alpha$. For example, we have $c^{tmax}=(0.0196,~0.088)$ for $\alpha=(1,~100)$. Nonetheless as already observed for other inflationary models, the maximum value of the $c$ appears from the minimum value of the reheating temperature. 
  For a fixed value of $\alpha=1$, maximum values turn out to be $c^{max}=(0.025,0.0216)$ for effective equation of state $\omega_{re}=(0,\frac{1}{6})$.
  By using the above Eq.\ref{alpha1}, we can interpret that any particular values of $c$ ($c^{tmax}<c<c^{max}$) and $\alpha$, maximum allowed 
  values of temperature is different for a different equation of state $\omega_{re}$. For $(\alpha,c)=(1,0.02)$, the maximum allowed values of 
  reheating temperature are $6.9\times10^{13} ~GeV$ and $5.3\times10^{11}~GeV$ for $\omega_{re}=(0,\frac{1}{6})$. 
  In the 
same way, the maximum and minimum value of the reheating temperature from observation 
imposes a restriction on the allowed values of $c$. After combining Eq.\ref{alpha1} and Eq.\ref{alpha1'} one can find
\bea
(c~c')^2~\leq \frac{128 ~n^4}{27 ~\alpha^3}~\frac{\left(e^{\sqrt{\frac{2}{3\alpha}}\frac{\phi}{M_p}}-2n\right)^2}{\left(e^{\sqrt{\frac{2}{3\alpha}}\frac{\phi}{M_p}}-1\right)^6}~~.
\eea
By using above equation we are able to find out the resultant allowed parameter space of $c$ and $c'$ as shown in the last plot of Fig.\ref{compare}.
\\
  At a fixed values of $M_X=10^3$ $GeV$ and $\alpha=1$, the lower limit of the 
 annihilation cross-section $\langle\sigma v \rangle ^{lower}$ is restricted by $c$, for instance, $\langle \sigma v \rangle ^{lower}\simeq 4.2\times 10^{-42}$ 
 for $c=0.0196$. So swampland conjecture constant $c$ restraint on the annihilation cross section $\langle \sigma v\rangle$ (fig.\ref{compare}) same way as shown in the chaotic 
 and natural inflation model, the only difference is that here the $c$ values are quite lower. The lower limit of the dark-matter annihilaton cross section start increasing with increasing $c$, where we consider $c$ with in $c^{max}>c>c^{tmax}$. As an example, $\langle \sigma v \rangle ^{lower}\simeq (1.9\times10^{-40},1.4\times10^{-36})~ GeV^{-2}$ 
 for $c=(0.020, 0.021)$.
\subsection{Supergravity inspired minimal plateau model \cite{pankaj}}
\begin{figure}[t!]
 	\begin{center}
 		\includegraphics[width=005.41cm,height=04.1cm]{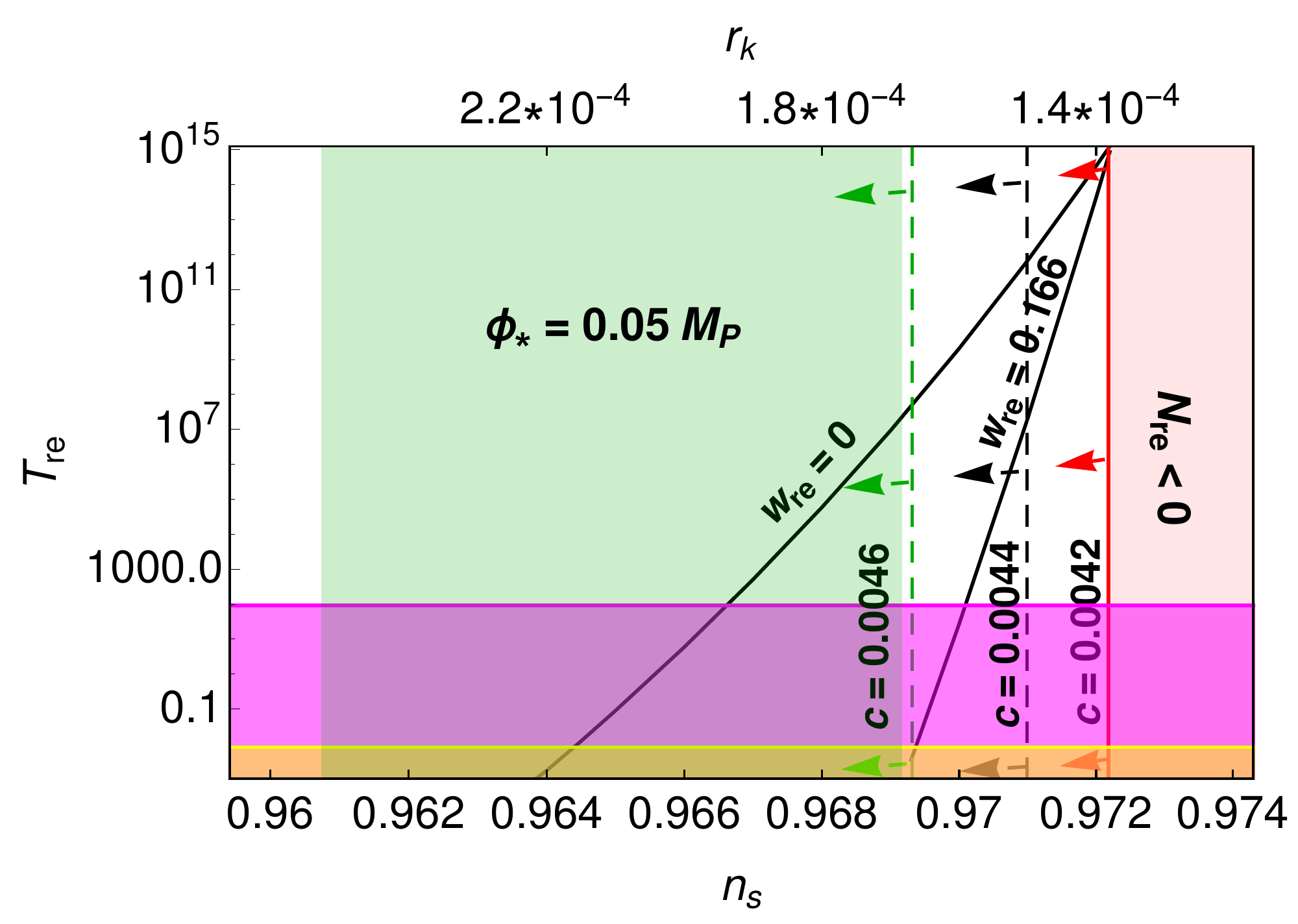}
 		\includegraphics[width=005.41cm,height=04.1cm]{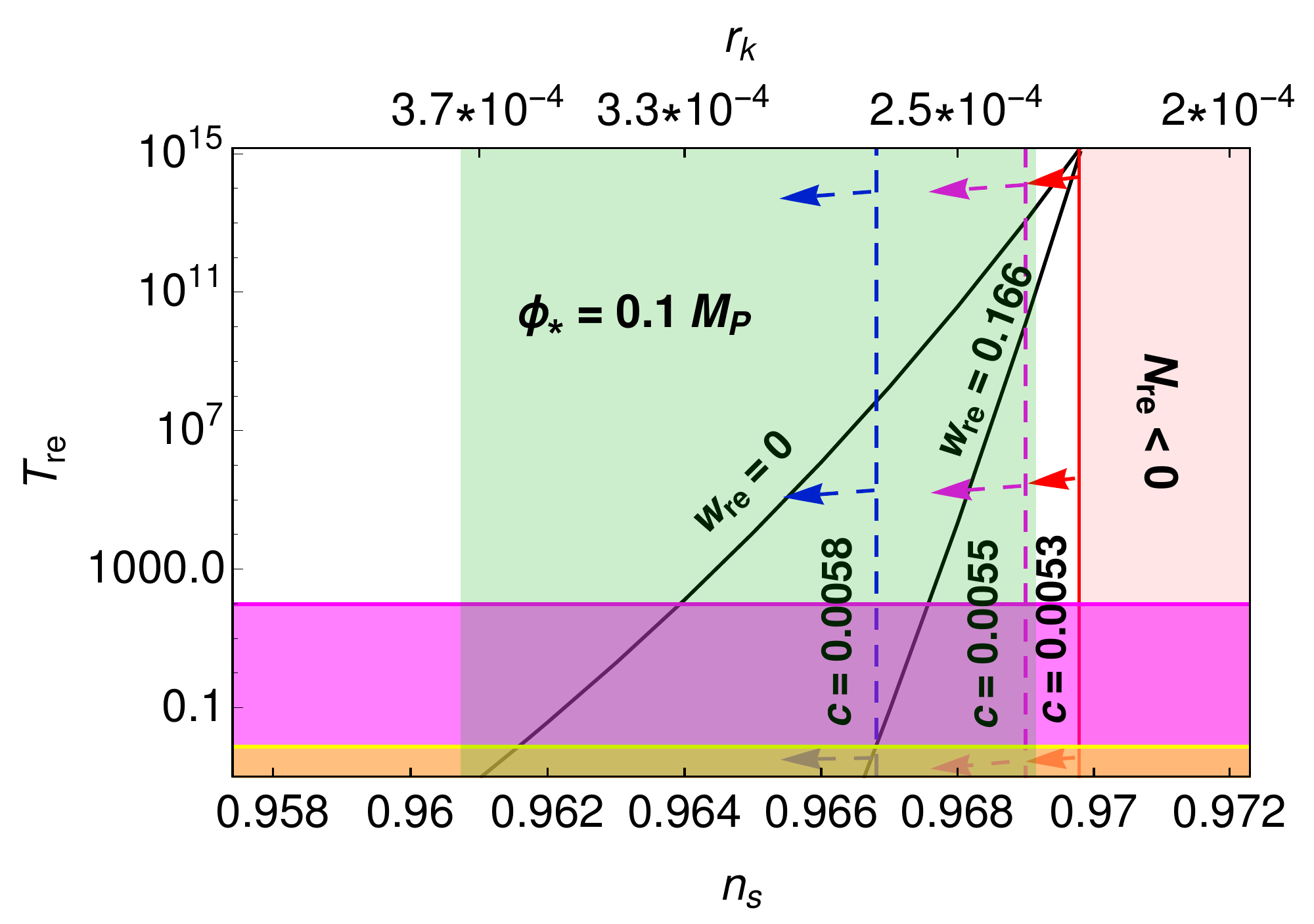}
 		\includegraphics[width=005.41cm,height=04.1cm]{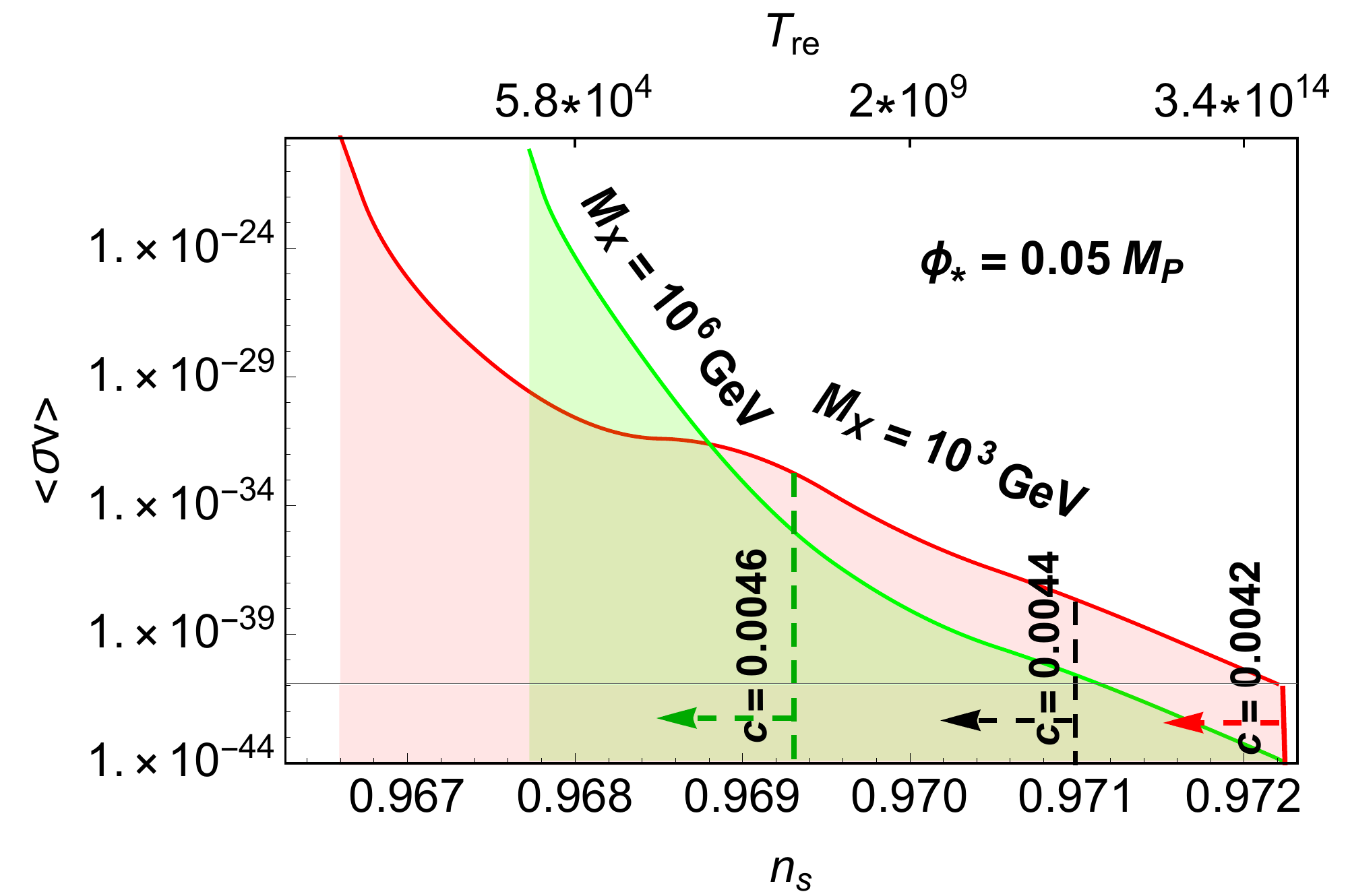}
 		\includegraphics[width=005.41cm,height=04.1cm]{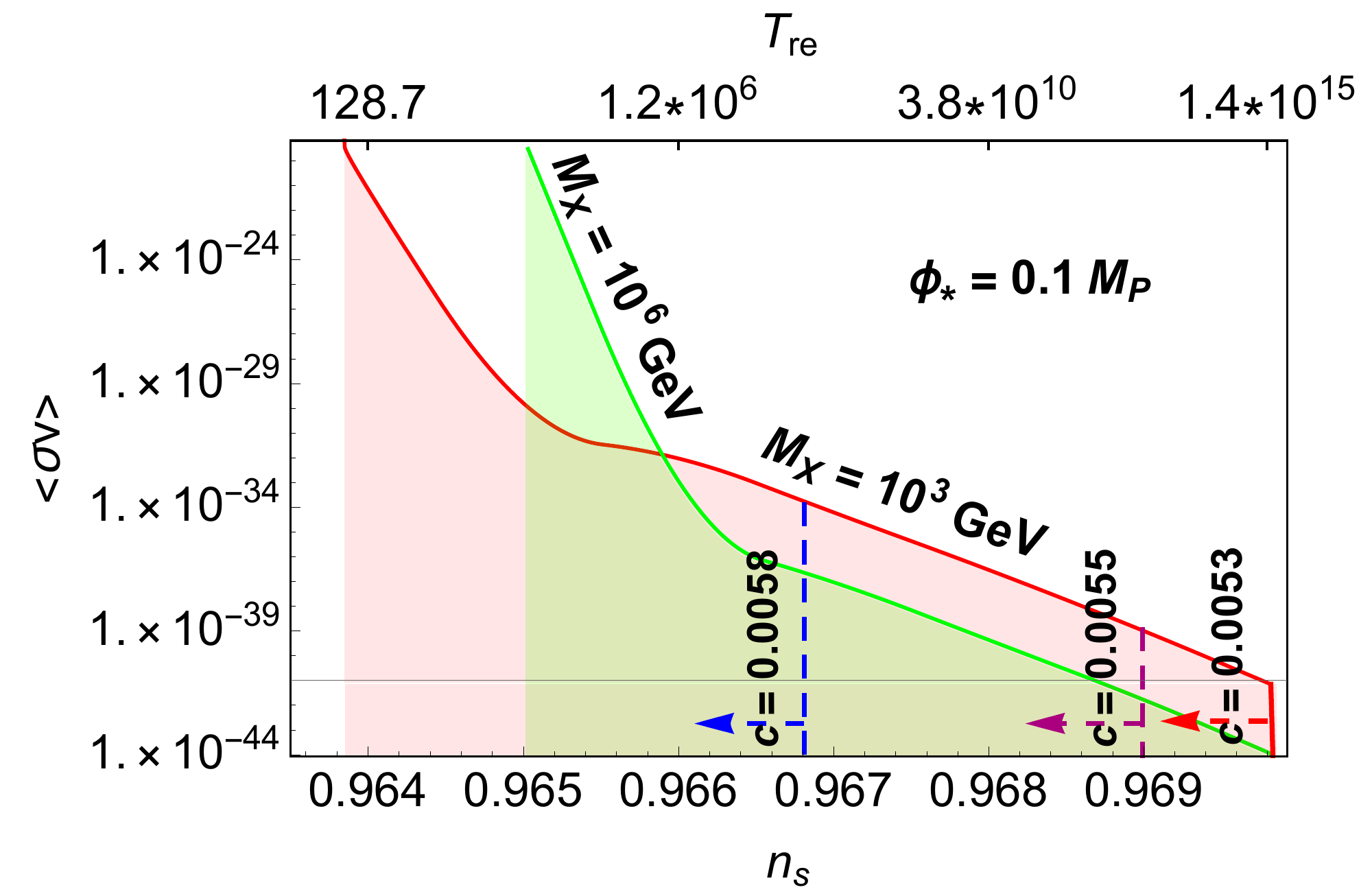}
 		\includegraphics[width=005.4cm,height=03.6cm]{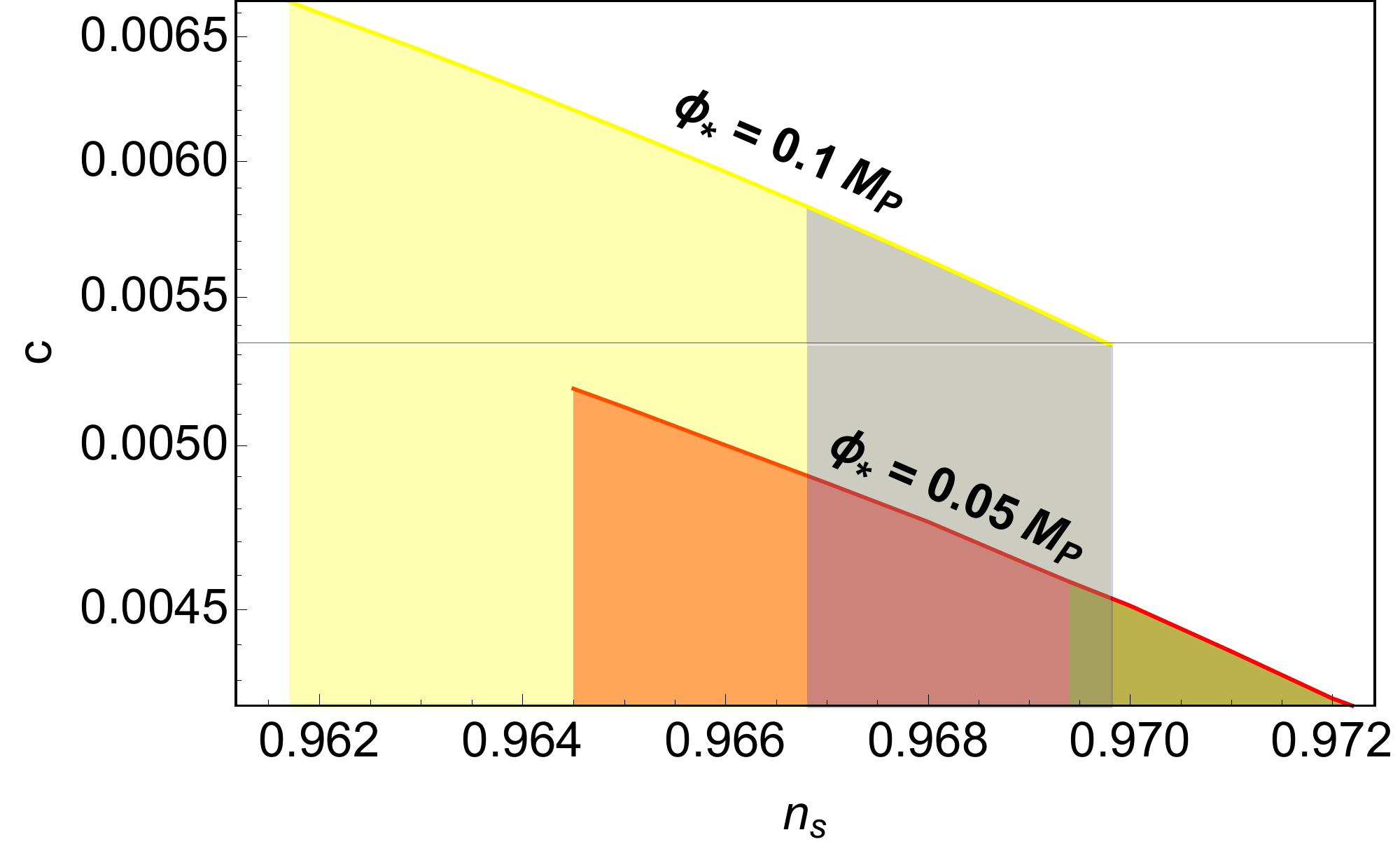}
 		\includegraphics[width=005.4cm,height=03.6cm]{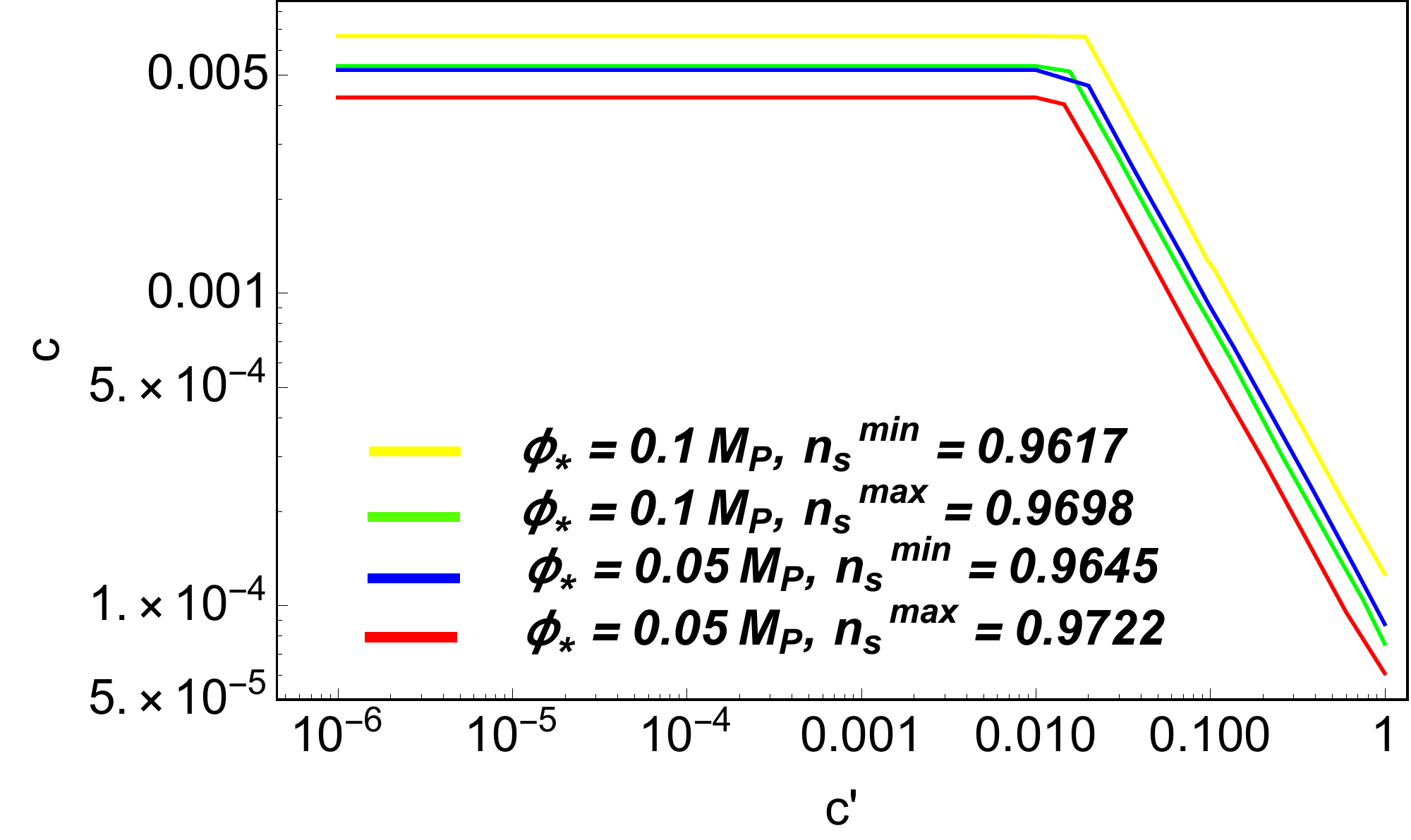}
 		\caption{\scriptsize All plots are same as in the previous fig.\ref{chaotic}. The main inequality is that here we have plotted for supergravity inflation model 
 		for $\phi_*=(0.05, 0.1)M_p$. In the lower middle plot, the region under 
 		 the yellow solid line is for ($\omega_{re}, \phi_*)=(0, 0.1~M_p)$ and dark blue region is for ($\omega_{re}, \phi_*)=(\frac{1}{6},0.1~M_p)$. In the 
 		 same plot, the region under 
 		 the red solid line is for ($\omega_{re}, \phi_*)=(0,0.1~M_p)$ and dark green region is for ($\omega_{re}, \phi_*)=(\frac{1}{6}, 0.1~M_p)$.}
 		\label{sg10}
 	\end{center}
 \end{figure}

In this section we will consider a special class of supergravity inspired inflation potential,
\bea \label{Supergravity} 
 V_{SG}(\phi)=\frac{m^{4-n}\phi^n}{e^{-\frac{\phi^2}{2 M_p^2}}+\left(\frac{\phi}{\phi_*}\right)^n}~~.
\eea
The shape of the potential depends on the mass scale $\phi_*$ and all other parameters are same as in the previous inflationary model. One of the striking features of this model is that unlike axion and $\alpha$-attractor models, it fits well within PLANCK data for all possible values of $\phi^*$ from super to sub-Planckian value. Our initial motivation was to figure out if with increasing $\phi^*$ the value of $(c,c')$ increases towards unity. However we did not find such solutions. None the less for our numerical computation, we have chosen two values of $\phi_*=(0.05,0.1) M_p$. At the point of instant reheating $N_{re}\simeq0$, maximum scalar spectral index assumes $n_s^{max}\simeq(0.0042,0.0053)$ for the aforementioned two values of  $\phi_*=(0.05,0.1)M_p$ respectively. 
Using the swampland conjecture the condition between the potential $V_{SG}(\phi)$ and the spectral index, $n_s$ (also reheating temperature) satisfis the following inequalities 
 \bea\label{sg1}
 M_p~\frac{V_{SG}'}{V_{SG}}=\frac{2M_p^2\phi_*^2+\phi_*^2\phi^2}{M_p\phi_*^2\phi+e^{\frac{\phi^2}{2M_p^2}}M_p\phi^3} \geq c~~,
 \eea
 
 \bea
 M_p^2~\frac{V_{SG}''}{V_{SG}}\leq -c'~~,
 \eea
 where
 \bea\label{sg2}
 M_p^2~\frac{V_{SG}''}{V_{SG}}=\frac{\phi_*^2\left(2M_p^4\left(\phi_*^2-3e^{\frac{\phi^2}{2M_p^2}}\phi^2\right)+M_p^2\phi^2\left(5\phi_*^2-3e^{\frac{\phi^2}{2M_p^2}}\phi^2\right)+\phi^4\left(\phi_*^2-e^{\frac{\phi^2}{2M_p^2}}\phi^2\right)\right)}{M_p^2\phi^2\left(\phi_*^2+e^{\frac{\phi^2}{2M_p^2}}\phi^2\right)^2}~~.
 \eea
The parameter $c$ assumes its maximum possible value $c^{tmax}$, at the point of instant reheating for  different equation of state with fixed $\phi_*$. For example, $c^{tmax}=0.0053$ for $\phi_*=0.1 M_p$ and it decreases with $\phi_*$ very slowly. Although maximum values of $c$ ($c^{max}$) are different for different values of equation of state but changes with different $\phi_*=0.1 M_p$, as $c^{max}=(0.0066,0.0058)$ with the following effective equation of state $\omega_{re}=(0,\frac{1}{6})$ respectively. Similar to other models discussed before, any particular value of $c$ between $c^{max}>c>c^{tmax}$ provides a upper limit on the reheating temperature for a given value of $\phi_*$ and $\omega_{re}$ as shown in the first and second plots of Fig.\ref{sg10}. For example maximum allowed values of the reheating temperature turns out to be $T_{re}^{max}=(1.1\times10^{13},1.3\times10^{10})~GeV$ for two different effective equation of state $\omega_{re}$=$(0,\frac{1}{6})$ considering fixed values of ($\phi_*,c$)=$(0.1 M_p,0.0055)$. Inversely, we can say that observational limit of reheating temperature imposes a restriction on the allowed free parameter space of the swampland conjecture ($c,c'$). In order to explain that, as has been done for previous model, one combines Eq.\ref{sg1} and Eq.\ref{sg2}, and the inequality turns out to be,
\bea\label{sg4}
(c ~c')^2\leq \frac{\phi_*^8\left(2M_p^2+\phi^2\right)^2\left(2M_p^4\left(\phi_*^2-3e^{\frac{\phi^2}{2M_p^2}}\phi^2\right)+M_p^2\phi^2\left(5\phi_*^2-3e^{\frac{\phi^2}{2M_p^2}}\phi^2\right)\right)^2}{M_p^6~\phi^6\left(\phi_*^2+e^{\frac{\phi^2}{2M_p^2}}\phi^2\right)^6}\nonumber\\
+\frac{\phi_*^8\left(2M_p^2+\phi^2\right)^2\left(\phi^4\left(\phi_*^2-e^{\frac{\phi^2}{2M_p^2}}\phi^2\right)\right)^2}{M_p^6~\phi^6\left(\phi_*^2+e^{\frac{\phi^2}{2M_p^2}}\phi^2\right)^6}~~.
\eea
The resulting constraints on ($c,c'$) is displayed in the last plot of the Fig.\ref{sg10}.\\
Now we will discuss the effect of the swampland conjecture on the dark-matter parameter space. As already mentioned in the previous inflationary model, the lower limit of the annihilation cross-section modified by different values of $c$ once we fixed dark-matter mass $M_X$ and mass scale of the potential $\phi_*$. For $M_X=10^3~GeV$ and $\phi_*=0.1 M_p$, $\langle \sigma v \rangle ^{lower}\simeq (8\times 10^{-42},1\times10^{-39},1.9\times10^{-34})~ GeV^{-2}$ 
 for $c=(0.0053,0.0055,0.0058)$. Therefore, most importantly the annihilation cross-section of the  dark-matter becomes more restricted with increasing $c$ within $c^{max}>c>c^{tmax}$. 
 \section{Summary and discussion:}
 Swampland conjecture has gained significant attention mainly because of its potential to validate or invalidate large number of low energy effective theories proposed in diverse physics problems. It mainly deals with a scalar field and its possible nature of the potential which is conjectured to follow certain constraints. Two parameters $(c,c')$ are conjectured to be of the order unity such that the field theory under consideration can have consistent ultraviolet completion when minimally coupled with gravity.
One of the best candidates scalar field is inflaton which has been proved to be successful in explaining large volume of cosmological observations. However, it turned out that more successful a inflation scenario is more incompatible with the Swampland conjecture it becomes. In the present paper instead of taking swampland parameters to be of order unity, we considered them free and analyze its impact on the other cosmological parameters with special emphasis on the reheating phase.  
Based on our analysis so far let us try to point out the main outcomes. We have considered four different types of inflaton potential and studied the consequence of swampland conjectures on those and constrain the parameter space. However, we must say that the conventional slow roll potential is very much constrained by the conjecture. For all the models under consideration what we found is that the possible values of $c$ is always less than unity. The maximum possible value one could get $c\simeq 10^{-1}$ is for chaotic inflation which predicts higher value of tensor to scalar ratio. Moreover we studied other models, such as axion, $\alpha$-attractor, supergravity inspired inflation, which are consistent with PLANCK data, and maximum possible value of $c$ turns out to even smaller for those model. This is intimately connected with smaller prediction of tensor to scalar ratio $r$. However, this has already been observed before. Our focus in this paper was more on the impact of this inflationary constraints of $(c,c')$ on the reheating phase and dark matter phenomenology. In Fig.\ref{chaotic}, \ref{NATURAL}, \ref{compare}, and \ref{sg10} we have considered different cosmological parameters such as $(n_s, T_{re}, \langle \sigma v\rangle)$, and studied their interdependence and constraints from the swampland conjecture. In the usual reheating constraint analysis, the reheating temperature varies widely within $2\sigma$ range of $n_s$, and consequently so does the dark matter cross-section $\langle \sigma v\rangle$ given a dark matter mass. Since swampland conjecture is an inequality, for the model under considerations it provides us an upper bound on $T_{re}$ and lower bound on $\langle \sigma v\rangle$ for a given dark matter mass $M_X$ and the swampland parameter $c$ . However since reheating temperature is already constrained by the BBN, for all the models we have an associated maximum value of swampland parameter $c^{max}$ as shown in the Table.\ref{table1}, where we considered a natural value of the effective equation of state $\omega_{re}=0$.

\begin{table}[t!]
	\caption{Models and their associated $c^{max}$ value}
  \begin{tabular}{|p{1.00cm}|p{1.00cm}|p{1.75cm}|p{1.750cm}|p{1.75cm}| p{1.75cm} |p{2cm} |p{1.85cm}|}
\hline
\hline
 &chaotic &\multicolumn{2}{c|}{$\alpha$-attractor}    &\multicolumn{2}{c|}{Axion} &\multicolumn{2}{c|}{Supergravity}\\
\cline{3-8}
 && $\alpha=1$ &  $\alpha=100$  &  $f=10M_p$  & $f=50M_p$&$\phi_* =0.05M_p$&$\phi _*=0.1M_p$ \\
\hline
\hline
 \specialcell{$c^{max}$}   &0.1448 & 0.025   & 0.0104& 0.128& 0.144& 0.0052& 0.0053\\
\hline
  \end{tabular}
  \label{table1}
\end{table}
Considering different class of inflationary models, one of our important observation is the existence of maximum possible reheating temperature $T_{re}^{max} \simeq 10^{15} $ GeV irrespective of the models. Importantly, however, the associated swampland parameter $c= c^{tmax}$ are depending upon the models and their parameters. 
In the same way even more interesting fact is that for a fixed value of $c$ within  $(c^{max}, c^{tmax})$, there exists associated maximum value of reheating temperature $T_{re}^{max}(c)$ and consequently  
minimum value of $\langle \sigma v\rangle^{lower}(c)$ for fixed dark matter mass.
For example, if we consider $c=0.13$,  $T_{re}^{max}(c)=(8.4\times10^{12},1.1\times 10^{12})$ $GeV$ and $\langle\sigma v \rangle ^{lower}(c)\simeq (1.5\times 10^{-39},1.4\times10^{-38})~ GeV^{-2}$ for chaotic and natural inflation model for $f=50M_p$ respectively. In $\alpha$-attractor model $T_{re}^{max}(c)=6.9\times10^{13}~ GeV$ and $\langle\sigma v \rangle ^{lower}(c)\simeq 1.9\times 10^{-40}~ GeV^{-2}$ for $c=0.02$ with $\alpha=1$, which is the well known Higgs/Starobinsky inflation model. Finally the supergravity inspired minimal potential which agrees very well with the PLANCK data at all values of $\phi_*$ turned out to set maximum possible possible of $c$ to be $c\simeq 10^{-2}$. Finally for all the models under consideration the value of $c$ turned out to be bounded within $(0,c^{max})$, and that of $c'$ is within $(0,1)$. 
 
 \hspace{0.5cm}

 \end{document}